\documentstyle[preprint,tighten,aps]{revtex}

\begin{document}
\preprint{\begin{minipage}{2in}\begin{flushright}
  UCSD/PTH/97-05 \\ UK/97-02 \\ hep-lat/9704011 \end{flushright}
  \end{minipage}}
\input epsf
\draft
\title{Efficient glueball simulations on anisotropic lattices}
\author{Colin~J.~Morningstar}
\address{Dept.~of Physics, University of California at San Diego,
  La Jolla, California 92093-0319}
\author{Mike Peardon}
\address{Department of
Physics and Astronomy, University of
Kentucky, Lexington, KY 40506-0055}
\date{April 22, 1997}
\maketitle
\begin{abstract}
Monte Carlo results for the low-lying glueball spectrum using an improved,
anisotropic action are presented.  Ten simulations at lattice spacings
ranging from 0.2 to 0.4 fm and two different anisotropies have
been performed in order demonstrate the advantages of using coarse,
anisotropic lattices to calculate glueball masses.  Our determinations
of the tensor ($2^{++}$) and pseudovector ($1^{+-}$) glueball masses are
more accurate than previous Wilson action calculations.
\end{abstract}
\pacs{PACS number(s): 11.15.Ha, 12.38.Gc, 12.39.Mk}
%
%\narrowtext

\section{Introduction}
\label{sec:intro}

Numerical simulations of gluons on a space-time lattice provide at present
the most reliable means of studying glueballs.  Glueball correlation
functions are, however, notoriously difficult quantities to measure in
Monte Carlo simulations: since the masses of these states are rather high
and their creation operators have large vacuum fluctuations, the
signal-to-noise ratio falls extremely rapidly as the separation between the
source and sink is increased.  Because of this, reliable studies of glueballs 
on fine lattices have required prohibitively large computer resources.  Thus, 
the development of more efficient simulation techniques in lattice QCD is 
crucial to establishing a detailed description of glueballs and their 
interactions.

The objective of this work is to examine the effectiveness
of using an improved, anisotropic lattice action to reduce the
computational effort needed to determine the glueball spectrum
in quenched QCD.  Improved actions allow access to continuum physics
on coarser lattices than possible using the simple Wilson
discretization.  Coarse lattice simulations are
more efficient for several reasons: for a given physical
volume, much fewer lattice sites are needed; the alleviation of critical
slowing down permits the faster generation of statistically
independent gauge-field configurations; glueball operator smearing
is faster due to the decreased number of links and a decrease in
the number of smearing iterations required; glueball wavefunctions
extend over much fewer lattice sites on a coarse lattice, making
the variational technique far more effective when using a feasible
number (a dozen or so) of basis operators.

However, for glueball mass calculations, the coarseness of the
temporal lattice spacing is a severe drawback. As the masses in lattice units
of the states of interest are so large, the number of correlator time intervals
which can be measured is reduced greatly \cite{UsLAT95}.
A straightforward solution to this problem which preserves the
computational advantages of coarse lattices\cite{colinLAT96,usLAT96,Alford96}
is to make use of anisotropic lattices in which the temporal spacing
is much smaller than that in the spatial directions.  This enables us to
exploit the enhanced signal-to-noise of the correlation functions at smaller
temporal separations.  A natural scale for the temporal lattice spacing
should be the inverse of the energy of the states of interest; thus, for
glueballs, a temporal cut-off larger than 1.5 GeV allows resolution from
accessible statistics of the correlator over a few time-slices.  Meanwhile,
the scale for the spatial lattice should be set by the size of the
wavefunction of the state; a spatial grid separation in the range
$0.2 - 0.4$ fm would seem reasonable.

Since we propose to use lattices in which the temporal lattice spacing
is small, improvement of the discretization in this direction is not
needed.  Thus, a lattice action
which couples only nearest-neighbor time-slices can be used.  The
transfer matrix corresponding to such an action is Hermitian and
positive definite; all of our effective masses must converge to
their plateau values monotonically from above.  This ensures the
validity of variational techniques which minimize the effective masses at
small temporal separations.  Such techniques are very effective in
diminishing the excited-state contributions to the glueball correlation
functions and are crucial for efficient extraction of ground-state masses.

In this paper, we demonstrate the increased efficiency of glueball
simulations using these actions on anisotropic lattices.
We present results for the masses of three of the
lighter SU(3) glueball states, the scalar ($0^{++}$), the tensor ($2^{++}$),
and the pseudovector ($1^{+-}$).  The masses of the first excited states
in the scalar and tensor channels were also examined.  Ten simulations
at lattice spacings ranging from 0.2 to 0.4 fm were performed, enabling
reliable extrapolations to the continuum limit (although the mass of the
scalar glueball was somewhat problematic).  The results are compared
to previous simulation data obtained using the Wilson action and we find that
more accurate determinations of the tensor and pseudovector glueball masses
have been achieved.  A comparison of efficiencies is also made.  Lastly, finite
volume effects are shown to be small.

The new action used in our simulations is described in
Sec.~\ref{sec:action}.  The details of the glueball simulations, including
the construction of the glueball operators, the generation of the gauge-field
configurations, and the analysis of the Monte Carlo data, are given in
Sec.~\ref{sec:simulation}.  The hadronic scale $r_0$ is used to relate
our results at different values of the coupling $\beta$ and the
aspect ratio $\xi$.  The determination of this scale in terms of the
lattice spacing using the static potential is outlined in
Sec.~\ref{sec:r0as}.  Sec.~\ref{sec:resdiscuss} contains our results
and discussion:  the glueball mass measurements are presented in
detail; finite volume effects are studied; extrapolations of the
masses at finite spacing to the continuum limit are undertaken; the
conversion of our results into physical units is described; and a
comparison of efficiencies with Wilson action simulations is made.
Our conclusions are given in Sec.~\ref{sec:conclusions}, along with
an outline of future work.

\section{An improved, anisotropic discretization of QCD}
\label{sec:action}

Our glueball mass determinations rely on numerical simulations
of glueballs on a Euclidean space-time lattice with spatial
and temporal spacings $a_s$ and $a_t$, respectively.
The improved gluonic action used in this study is given by
\cite{colinLAT96,Alford96}
\begin{equation}
S_{II}  = \beta\ \left\{ \frac{5}{3} \frac{\Omega_{\rm sp}}{\xi u_s^4}
    +\frac{4}{3}\frac{\xi \Omega_{\rm tp}}{u_s^2u_t^2}
     - \frac{1}{12}\frac{\Omega_{\rm sr}}{\xi u_s^6}
     - \frac{1}{12}\frac{\xi \Omega_{\rm str}}{u_s^4u_t^2}  \right\},
\label{eq:ImpAction}
\end{equation}
where $\beta=6/g^2$, $g$ is the QCD coupling, $u_s$ and $u_t$ are mean
link renormalization parameters, $\xi$ is the aspect ratio ($\xi=a_s/a_t$
at tree level in perturbation theory), and
$\Omega_C=\sum_C {\textstyle\frac{1}{3}}\mbox{Re Tr}(1-W_C)$,
with $W_C$ denoting the path-ordered product of link variables
along a closed contour $C$ on the lattice.
$\Omega_{\rm sp}$ includes the sum over all spatial plaquettes on the
lattice, $\Omega_{\rm tp}$ indicates the temporal plaquettes,
$\Omega_{\rm sr}$ denotes the product of link variables about planar
$2\times 1$ spatial rectangular loops, and $\Omega_{\rm str}$ refers
to the short temporal rectangles (one temporal link, two spatial).
Explicitly,
\begin{eqnarray}
\Omega_{\rm sp} &=& \sum_x\sum_{i>j} \textstyle{\frac{1}{3}}{\rm Re Tr}
 \left[1-U_i(x)U_j(x\!+\!\hat\imath)U^\dagger_i(x\!+\!\hat\jmath)U^\dagger_j(x)
 \right],\\
\Omega_{\rm tp} &=& \sum_x\sum_{i} \textstyle{\frac{1}{3}}{\rm Re Tr}
 \left[1-U_t(x)U_i(x\!+\!\hat t)U^\dagger_t(x\!+\!\hat\imath)U^\dagger_i(x)
 \right],\\
\Omega_{\rm sr} &=& \sum_x\sum_{i\neq j} \textstyle{\frac{1}{3}}{\rm Re Tr}
 \left[1-U_i(x)U_i(x\!+\!\hat\imath)U_j(x\!+\!2\hat\imath)
 U^\dagger_i(x\!+\!\hat\imath\!+\!\hat\jmath) U^\dagger_i(x\!+\!\hat\jmath)
 U^\dagger_j(x)\right],\\
\Omega_{\rm str} &=& \sum_x\sum_{i} \textstyle{\frac{1}{3}}{\rm Re Tr}
 \left[1-U_i(x)U_i(x\!+\!\hat\imath)U_t(x\!+\!2\hat\imath)
 U^\dagger_i(x\!+\!\hat\imath\!+\!\hat t) U^\dagger_i(x\!+\!\hat t)
 U^\dagger_t(x)\right],
\end{eqnarray}
where $x$ labels the sites of the lattice, $i,j$ are spatial
indices, and $U_\mu(x)$ is the parallel transport matrix in the gluon field
from site $x$ to $x\!+\!\hat\mu$.

This action, intended for use with $a_t \ll a_s$, has
$O(a^4_s,a^2_t,\alpha_s a_s^2)$ discretization errors.
The $O(a_t^2)$ errors can be removed by the addition
of counterterms which couple next-nearest-neighbor time-slices,
but this introduces spurious high-energy modes which can cause
considerable problems for our glueball simulations.  These unphysical
states appear in perturbation theory as additional poles in the
gluon propagator.  Their detrimental effects on the glueball
correlation functions have been previously demonstrated\cite{UsLAT95}.
Although these spurious states do not affect the asymptotic behavior
of the glueball correlators, they do appreciably change the correlators
at short temporal separations and can seriously hinder attempts to reduce
excited-state contamination to hasten the onset of asymptotic behavior.
Since our glueball mass measurements rely heavily on the reduction of such
excited-state contributions to the glueball correlation functions,
the use of an action which is free of spurious lattice modes is crucial.
The action given in Eq.~\ref{eq:ImpAction} couples only link variables on 
neighboring time-slices, which ensures that all of our effective masses 
converge to their plateau values monotonically from above and so validates the 
variational techniques employed.

It is now known that perturbation theory by itself does not reliably
determine the couplings in an improved action in lattice gauge theory.
Hence, the interaction strengths in $S_{I\!I}$ have been determined
using a judicious combination of perturbation theory and mean field
theory.  Mean field theory is introduced by separately renormalizing
the spatial and temporal link variables:
$U_j(x) \rightarrow U_j(x) / u_s$ and
$U_t(x) \rightarrow U_t(x) / u_t$, where $u_s$ and $u_t$ denote the
renormalization factors for the spatial and temporal links,
respectively.  The mean-link parameters $u_t$ and $u_s$ are best
determined by guessing input values for use in the action, measuring
the mean links in Landau gauge in a simulation, then readjusting the
input values accordingly and tuning until the input values match
the measured values.  The determination of these renormalization factors
is described in more detail in Refs.~\cite{colinLAT96,Alford96}.
However, when $a_t$ is significantly smaller than $a_s$, we expect the
mean temporal link $u_t$ to be very close to unity since
$1-\langle\mbox{$\frac{1}{3}$}{\rm Tr} U_t\rangle \propto (a_t/a_s)^2$
in perturbation theory.  Hence, to simplify matters, we set $u_t = 1$.
We introduce further simplifications by using a convenient and
gauge-invariant definition for $u_s$ in terms of the mean spatial
plaquette given by $u_s=\langle \frac{1}{3}{\rm ReTr} P_{ss^\prime}
\rangle^{1/4}$, where $P_{ss^\prime}$ denotes the spatial plaquette.
This eliminates the need for gauge fixing, yields values for $u_s$
which differ from those found using the Landau gauge definition by
only a few per cent, and significantly speeds up the tuning process.

At finite coupling $g$, the anisotropy $a_s/a_t$ is renormalized away
from its input value $\xi$.  Measurements of this renormalization have
been made using the static potential extracted from correlations along
the different spatial and temporal axes of the lattice
\cite{colinLAT96,Alford96}.  Without mean-link improvement, this
renormalization can be as large as $30\%$.  When the action includes
mean-link corrections, this renormalization is found to be small,
typically a few per cent. We used $a_t/a_s=\xi$ in all of our calculations, 
accepting the small radiative corrections to the anisotropy as finite lattice 
spacing errors, which vanish in the continuum limit.

\section{Glueball simulation details}
\label{sec:simulation}

Glueballs may be labeled by their total (integral) spin $J$ and their
symmetries under spatial inversion and charge conjugation.  However,
on a cubic lattice, glueballs are characterized by their transformation
properties under the cubic point group, combined with parity and charge
conjugation operations. The cubic group, $O_h$, has 24 elements that fall
into five conjugacy classes, and thus, the dimensions of the irreducible
representations (irreps) are 1, 1, 2, 3 and 3.  These irreps are labeled
$A_1$, $A_2$, $E$, $T_1$, and $T_2$, respectively.  Including parity and
charge conjugation symmetry operations, there are 20 irreps (labeled by
$J^{PC}$, where $J$ now denotes an irrep of $O_h$).  In this study, four of the
irreps which generate light ($<3$ GeV) glueball states were simulated:
the $A_1^{++}$, $E^{++}$, $T_2^{++}$, and $T_1^{+-}$ channels.
Of particular interest are the $E^{++}$ and $T_2^{++}$ irreps whose combined
five rows correspond to the five polarization states of the tensor ($2^{++}$)
glueball which become degenerate as continuum rotational invariance is
restored. This then gives information on the magnitude of lattice artifacts at
finite cut-off.

The mass of a glueball, $G$, having a given $J^{PC}$
can be extracted from the large-$t$ behavior of a
correlation function $C(t)=\langle 0\vert \bar{\Phi}^{(R)\dagger}(t)
\ \bar{\Phi}^{(R)}(0)\vert 0\rangle$, where $R$ denotes the lattice
irrep corresponding to the $J^{PC}$ of interest and $\bar{\Phi}^{(R)}(t)
=\Phi^{(R)}(t)-\langle 0\vert\Phi^{(R)}(t)\vert 0\rangle$ is a gauge-invariant,
translationally-invariant, vacuum-subtracted operator capable of creating
a glueball out of the QCD vacuum $\vert 0\rangle$.  As the temporal
separation $t$ becomes large, this correlator tends to a single decaying
exponential $\lim_{t\rightarrow \infty}C(t) = Z \exp(-m_G t)$, where $m_G$ is
the mass of the lowest-lying glueball which can be created by the operator
$\bar{\Phi}^{(R)}(t)$.  In order to extract $m_G$, the correlator $C(t)$ must
be determined for $t$ sufficiently large that $C(t)$ is
well approximated by its asymptotic form.  However, the signal-to-noise
ratio in any Monte Carlo determination of $C(t)$ falls exponentially fast
with respect to $t$.  Thus, it is crucial to use a glueball operator
for which $C(t)$ attains its asymptotic form as quickly as possible.
If $\vert G\rangle$ denotes the glueball state of interest, this means
that we must choose an operator for which the overlap
$\langle G\vert \bar\Phi^{(R)}(t)\vert 0\rangle / [\langle G\vert G\rangle
\langle 0\vert \bar\Phi^{(R)\dagger}(t)\bar\Phi^{(R)}(t)\vert 0\rangle]^{1/2}$
is as near to unity as possible.  For such an operator, the
signal-to-noise ratio is also optimal\cite{Brandstaeter}.

In order to construct such operators, we exploited the smearing
\cite{APESmear,TeperFuzz} and variational techniques which have been
used with success in earlier Wilson action simulations.  In each of
the $J^{PC}$ channels of interest, glueball operators were constructed
on each time-slice in a sequence of three steps. First, smeared links
$U^s_j(x)$ and fuzzy superlinks $U^f_j(x)$ were formed. Secondly, a set
of basic operators $\phi^{(R)}_{\alpha}(t)$ were constructed using linear
combinations of gauge-invariant, path-ordered products of the $U^s_j(x)$
and $U^f_j(x)$ matrices about various closed spatial loops; each such
linear combination was designed to be invariant under spatial translations
and to transform irreducibly under the symmetry operations of the cubic
point group according to the irrep of interest.  Lastly, the glueball
operators $\Phi^{(R)}(t)$ were formed from linear combinations of the basic
operators, $\Phi^{(R)}(t)=\sum_\alpha v_\alpha^{(R)} \phi^{(R)}_{\alpha}(t)$,
where the coefficients $v^{(R)}_\alpha$ were determined using the
variational method.  Each of these three steps is described below.

Operators constructed out of smeared links and fuzzy superlinks
have dramatically reduced mixings with the high frequency modes of
the theory.  Thus, the use of spatially-smoothed links is an important
part of reducing excited-state contamination in the glueball correlation
functions.  Two smoothing procedures were used: a single-link procedure
and a double-link procedure.  In the single-link procedure, every spatial
link $U_j(x)$ on the lattice is replaced by itself plus a sum of
its four neighboring (spatial) staples, projected back into SU(3):
\begin{equation} %%%%% SMEARING EQUATION
U^s_j(x) = {\cal P}_{SU(3)}\ \biggl\{ U_j(x)
 + \lambda_s \sum_{\pm(k\neq j)}
 U_k(x)\ U_j(x\!+\!\hat k)\ U_k^\dagger(x\!+\!\hat\jmath)
\biggr\},
\end{equation}
where ${\cal P}_{SU(3)}$ denotes the projection \cite{su3project}
into SU(3).  Here, we denote this mapping of the spatial link
matrices into the smeared link variables by $s_{\lambda_s}$.  In
the double-link procedure, new superlinks $U^f$ of length $2a_s$ are
built using neighboring staples which connect sites separated by a
distance twice that of the length of the source link variables:
\begin{eqnarray}
U^f_j(x)  & = & {\cal P}_{SU(3)}\ \biggl\{ U_j(x)\ U_j(x+\hat\jmath)
   \nonumber \\
          & + & \lambda_f \sum_{\pm(k\neq j)}
U_k(x)\ U_j(x\!+\!\hat k)\ U_j(x\!+\!\hat\jmath+\!\hat k)\
U_k^\dagger(x\!+\!2\hat\jmath)
\biggr\},
\end{eqnarray}
and we denote this mapping by $f_{\lambda_f}$.  Both procedures can
be applied recursively; smeared links can be smeared again and fuzzy
links of increasing length $2a_s,\ 4a_s,\ 8a_s \dots$ can be constructed.
A {\em smoothing scheme}, ${\cal S}$, is defined as a composition of
single-link mappings and double-link mappings.  Six different smoothing
schemes were used.  The simplest scheme used was the composition of two
single-link smearings: ${\cal S}_1 = s_{\lambda_s} \circ s_{\lambda_s}$.
To simplify notation, we write this as ${\cal S}_1 = s_{\lambda_s}^2$.
We also used the compositions of four and six single-link mappings:
${\cal S}_2 = s_{\lambda_s}^4$ and ${\cal S}_3 = s_{\lambda_s}^6$.
In the other three smoothing schemes, the application of several
single-link smearings, followed by one final iteration of double-link
fuzzing was used:
${\cal S}_4 = f_{\lambda_f} \circ s_{\lambda_s}^2$,
${\cal S}_5 = f_{\lambda_f} \circ s_{\lambda_s}^4$, and
${\cal S}_6 = f_{\lambda_f} \circ s_{\lambda_s}^6$.
Only one iteration of the fuzzing procedure which results in links
connecting sites separated by $2a_s$ was found to be useful for the
range of coarse $a_s$ values explored here.  For the finer lattices
($\beta=2.4, \xi=5$ and $\beta=2.6, \xi=3$), an extra four initial
iterations of single-link smearing were used in all six smoothing
schemes to enhance ground-state overlap.  To simplify matters,
the same values for the two parameters $\lambda_s$ and $\lambda_f$ were
used in all smearing and fuzzing iterations.  These values were chosen
to minimize excited-state contamination in the glueball correlation
functions.  A crude optimization was done in a set of low statistics
runs and the optimal values $\lambda_s=0.1$ and $\lambda_f=0.5$
were then used in all the glueball simulations.

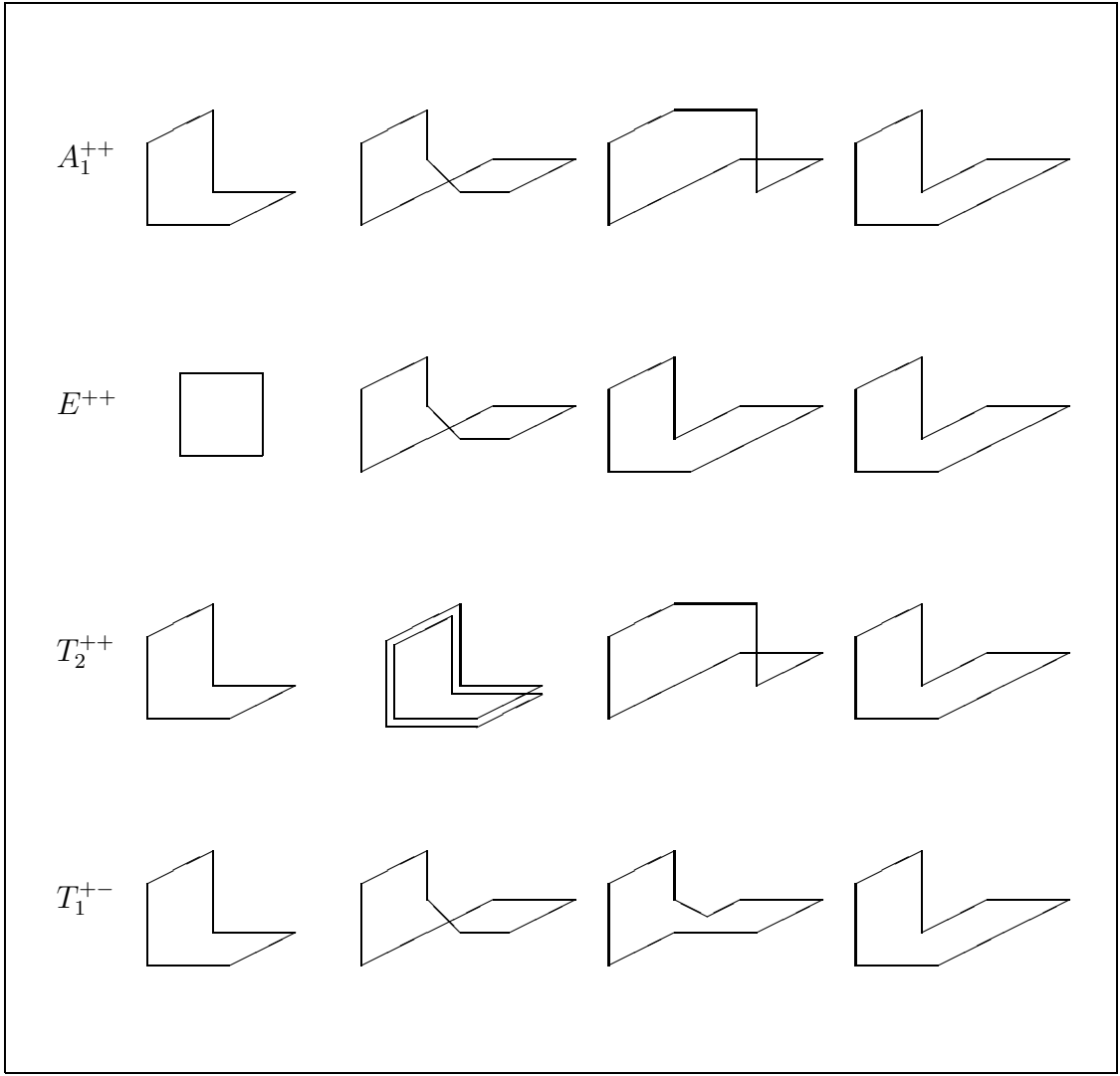
\begin{figure}
\begin{center}
\setlength{\unitlength}{0.11cm}
\begin{picture}(135,140)
\put(0,0){\line(1,0){135}}
\put(0,0){\line(0,1){130}}
\put(135,0){\line(0,1){130}}
\put(0,130){\line(1,0){135}}
\put(0,100){
 \begin{picture}(10,20)
 \put(5,10){$A_1^{++}$}
 \end{picture}}
\put(10,100){
 \begin{picture}(30,20)
 \put( 14.00, 7.00){\line( 1, 0){10}}
 \put( 16.00,  3.00){\line( 2, 1){ 8}}
 \put(  6.00,  3.00){\line( 1, 0){10}}
 \put(  6.00,  3.00){\line( 0, 1){10}}
 \put(  6.00, 13.00){\line( 2, 1){ 8}}
 \put( 14.00,  7.00){\line( 0, 1){10}}
 \end{picture}}
\put(40,100){
 \begin{picture}(30,20)
 \put( 10.00,  7.00){\line( 2, 1){ 8}}
 \put( 18.00, 11.00){\line( 1, 0){10}}
 \put( 20.00,  7.00){\line( 2, 1){ 8}}
 \put( 14.00,  7.00){\line( 1, 0){ 6}}
 \put(  2.00,  3.00){\line( 2, 1){ 8}}
 \put(  2.00,  3.00){\line( 0, 1){10}}
 \put(  2.00, 13.00){\line( 2, 1){ 8}}
 \put( 10.00, 11.00){\line( 0, 1){6}}
 \put( 10.00, 11.00){\line( 1,-1){4}}
 \end{picture}}
\put(70,100){
 \begin{picture}(30,20)
 \put( 10.00,  7.00){\line( 2, 1){ 8}}
 \put( 18.00, 11.00){\line( 1, 0){10}}
 \put( 20.00,  7.00){\line( 2, 1){ 8}}
 \put(  2.00,  3.00){\line( 2, 1){ 8}}
 \put( 10.00, 17.00){\line( 1, 0){10}}
 \put(  2.00,  3.00){\line( 0, 1){10}}
 \put(  2.00, 13.00){\line( 2, 1){ 8}}
 \put( 20.00,  7.00){\line( 0, 1){10}}
 \end{picture}}
\put(100,100){
 \begin{picture}(30,20)
 \put( 10.00,  7.00){\line( 2, 1){ 8}}
 \put( 18.00, 11.00){\line( 1, 0){10}}
 \put( 20.00,  7.00){\line( 2, 1){ 8}}
 \put( 12.00,  3.00){\line( 2, 1){ 8}}
 \put(  2.00,  3.00){\line( 1, 0){10}}
 \put(  2.00,  3.00){\line( 0, 1){10}}
 \put(  2.00, 13.00){\line( 2, 1){ 8}}
 \put( 10.00,  7.00){\line( 0, 1){10}}
 \end{picture}}
%%%%%%%%%%%%%%%%%%%%%%%%%%%%%%%%%%%%%%%%%%%%%%%%%%%%%%%%%%%%%%%
\put(0,70){
 \begin{picture}(10,20)
 \put(5,10){$E^{++}$}
 \end{picture}}
\put(10,70){
 \begin{picture}(30,20)
 \put( 10.00,  5.00){\line( 1, 0){10}}
 \put( 20.00,  5.00){\line( 0, 1){10}}
 \put( 10.00,  5.00){\line( 0, 1){10}}
 \put( 10.00, 15.00){\line( 1, 0){10}}
 \end{picture}}
\put(40,70){
 \begin{picture}(30,20)
 \put( 10.00,  7.00){\line( 2, 1){ 8}}
 \put( 18.00, 11.00){\line( 1, 0){10}}
 \put( 20.00,  7.00){\line( 2, 1){ 8}}
 \put( 14.00,  7.00){\line( 1, 0){ 6}}
 \put(  2.00,  3.00){\line( 2, 1){ 8}}
 \put(  2.00,  3.00){\line( 0, 1){10}}
 \put(  2.00, 13.00){\line( 2, 1){ 8}}
 \put( 10.00, 11.00){\line( 0, 1){6}}
 \put( 10.00, 11.00){\line( 1,-1){4}}
 \end{picture}}
\put(70,70){
 \begin{picture}(30,20)
 \put( 10.00,  7.00){\line( 2, 1){ 8}}
 \put( 18.00, 11.00){\line( 1, 0){10}}
 \put( 20.00,  7.00){\line( 2, 1){ 8}}
 \put( 12.00,  3.00){\line( 2, 1){ 8}}
 \put(  2.00,  3.00){\line( 1, 0){10}}
 \put(  2.00,  3.00){\line( 0, 1){10}}
 \put(  2.00, 13.00){\line( 2, 1){ 8}}
 \put( 10.00,  7.00){\line( 0, 1){10}}
 \end{picture}}
\put(100,70){
 \begin{picture}(30,20)
 \put( 10.00,  7.00){\line( 2, 1){ 8}}
 \put( 18.00, 11.00){\line( 1, 0){10}}
 \put( 20.00,  7.00){\line( 2, 1){ 8}}
 \put( 12.00,  3.00){\line( 2, 1){ 8}}
 \put(  2.00,  3.00){\line( 1, 0){10}}
 \put(  2.00,  3.00){\line( 0, 1){10}}
 \put(  2.00, 13.00){\line( 2, 1){ 8}}
 \put( 10.00,  7.00){\line( 0, 1){10}}
 \end{picture}}
\put(0,40){
 \begin{picture}(10,20)
 \put(5,10){$T_2^{++}$}
 \end{picture}}
\put(10,40){
 \begin{picture}(30,20)
 \put( 14.00, 7.00){\line( 1, 0){10}}
 \put( 16.00,  3.00){\line( 2, 1){ 8}}
 \put(  6.00,  3.00){\line( 1, 0){10}}
 \put(  6.00,  3.00){\line( 0, 1){10}}
 \put(  6.00, 13.00){\line( 2, 1){ 8}}
 \put( 14.00,  7.00){\line( 0, 1){10}}
 \end{picture}}
\put(40,40){
 \begin{picture}(30,20)
 \put( 14.00, 7.00){\line( 1, 0){10}}
 \put( 16.00,  3.00){\line( 2, 1){ 8}}
 \put(  6.00,  3.00){\line( 1, 0){10}}
 \put(  6.00,  3.00){\line( 0, 1){9}}
 \put(  5.00, 12.50){\line( 2, 1){9}}
 \put( 14.00,  7.00){\line( 0, 1){10}}
 \put( 13.00, 6.00){\line( 1, 0){11}}
 \put( 16.00,  2.00){\line( 2, 1){ 8}}
 \put(  5.00,  2.00){\line( 1, 0){11}}
 \put(  5.00,  2.00){\line( 0, 1){10.5}}
 \put(  6.00, 12.00){\line( 2, 1){ 7}}
 \put( 13.00,  6.00){\line( 0, 1){9.5}}
 \end{picture}}
\put(70,40){
 \begin{picture}(30,20)
 \put( 10.00,  7.00){\line( 2, 1){ 8}}
 \put( 18.00, 11.00){\line( 1, 0){10}}
 \put( 20.00,  7.00){\line( 2, 1){ 8}}
 \put(  2.00,  3.00){\line( 2, 1){ 8}}
 \put( 10.00, 17.00){\line( 1, 0){10}}
 \put(  2.00,  3.00){\line( 0, 1){10}}
 \put(  2.00, 13.00){\line( 2, 1){ 8}}
 \put( 20.00,  7.00){\line( 0, 1){10}}
 \end{picture}}
\put(100,40){
 \begin{picture}(30,20)
 \put( 10.00,  7.00){\line( 2, 1){ 8}}
 \put( 18.00, 11.00){\line( 1, 0){10}}
 \put( 20.00,  7.00){\line( 2, 1){ 8}}
 \put( 12.00,  3.00){\line( 2, 1){ 8}}
 \put(  2.00,  3.00){\line( 1, 0){10}}
 \put(  2.00,  3.00){\line( 0, 1){10}}
 \put(  2.00, 13.00){\line( 2, 1){ 8}}
 \put( 10.00,  7.00){\line( 0, 1){10}}
 \end{picture}}
%%%%%%%%%%%%%%%%%%%%%%%%%%%%%%%%%%%%%%%%%%%%%%%%%%%%%%%%%%%%%%%
\put(0,10){
 \begin{picture}(10,20)
 \put(5,10){$T_1^{+-}$}
 \end{picture}}
\put(10,10){
 \begin{picture}(30,20)
 \put( 14.00, 7.00){\line( 1, 0){10}}
 \put( 16.00,  3.00){\line( 2, 1){ 8}}
 \put(  6.00,  3.00){\line( 1, 0){10}}
 \put(  6.00,  3.00){\line( 0, 1){10}}
 \put(  6.00, 13.00){\line( 2, 1){ 8}}
 \put( 14.00,  7.00){\line( 0, 1){10}}
 \end{picture}}
\put(40,10){
 \begin{picture}(30,20)
 \put( 10.00,  7.00){\line( 2, 1){ 8}}
 \put( 18.00, 11.00){\line( 1, 0){10}}
 \put( 20.00,  7.00){\line( 2, 1){ 8}}
 \put( 14.00,  7.00){\line( 1, 0){ 6}}
 \put(  2.00,  3.00){\line( 2, 1){ 8}}
 \put(  2.00,  3.00){\line( 0, 1){10}}
 \put(  2.00, 13.00){\line( 2, 1){ 8}}
 \put( 10.00, 11.00){\line( 0, 1){6}}
 \put( 10.00, 11.00){\line( 1,-1){4}}
 \end{picture}}
\put(70,10){
 \begin{picture}(30,20)
 \put( 14.00,  9.00){\line( 2, 1){ 4}}
 \put( 18.00, 11.00){\line( 1, 0){10}}
 \put( 20.00,  7.00){\line( 2, 1){ 8}}
 \put( 10.00,  7.00){\line( 1, 0){10}}
 \put(  2.00,  3.00){\line( 2, 1){ 8}}
 \put(  2.00,  3.00){\line( 0, 1){10}}
 \put(  2.00, 13.00){\line( 2, 1){ 8}}
 \put( 10.00, 11.00){\line( 0, 1){6}}
 \put( 14.00,  9.00){\line(-2, 1){4}}
 \end{picture}}
\put(100,10){
 \begin{picture}(30,20)
 \put( 10.00,  7.00){\line( 2, 1){ 8}}
 \put( 18.00, 11.00){\line( 1, 0){10}}
 \put( 20.00,  7.00){\line( 2, 1){ 8}}
 \put( 12.00,  3.00){\line( 2, 1){ 8}}
 \put(  2.00,  3.00){\line( 1, 0){10}}
 \put(  2.00,  3.00){\line( 0, 1){10}}
 \put(  2.00, 13.00){\line( 2, 1){ 8}}
 \put( 10.00,  7.00){\line( 0, 1){10}}
 \end{picture}}
%%%%%%%%%%%%%%%%%%%%%%%%%%%%%%%%%%%%%%%%%%%%%%%%%%%%%%%%%%%%%%%
\end{picture}
\end{center}
\caption[glueops]{The four Wilson loop shapes in each channel used to
form the lattice glueball operators.  The complete set of 24 operators
was formed by computing linear combinations of each of these loops
rotated and translated across the lattice on six
different sets of smoothed links.  Where a loop shape occurs twice, it
is used in two different projections into the appropriate irreducible
representation.
\label{glueops}}
\end{figure}

The second step in the construction of our glueball operators was the
formation of a set of basic operators $\phi^{(R)}_{\alpha}(t)$ using
linear combinations of gauge-invariant, path-ordered products of the
$U^s_j(x)$ and $U^f_j(x)$ matrices about various closed spatial loops.
Combinations which were Hermitian, invariant under spatial translations,
and transformed irreducibly under the operations of the cubic
point group according to the irrep of interest were constructed.
For a more detailed exposition of this construction, see Ref.~\cite{Berg83}.
In each channel, a large set of prototypes were programmed, and
a short simulation was then performed to determine the coefficients of each
operator in the variational ground state.  In each channel, the four operators 
with the highest of these contributions were then chosen for use in the 
production runs.  The paths in this optimal set are illustrated in 
Fig.~\ref{glueops}.  In the glueball simulations, these Wilson loops were 
measured on the link variables from the six smoothing schemes, yielding a total 
of $N=24$ basic operators $\phi^{(R)}_{\alpha}(t)$ in each of the four channels.

Finally, $\Phi^{(R)}(t)$ was formed from a linear combination of the
basic operators, $\Phi^{(R)}(t)
 =\sum_{\alpha=1}^N v_\alpha^{(R)} \phi^{(R)}_{\alpha}(t)$.
The coefficients $v^{(R)}_\alpha$ were determined using the
variational method.  First, the $24 \times 24$ correlation matrix
was computed in the glueball simulations:
\begin{equation}
\tilde C_{\alpha\beta}(t) = \sum_\tau
            \langle 0\vert\bar{\phi}^{(R)}_{\alpha}(\tau\!+\!t) \:
            \bar{\phi}^{(R)}_{\beta}(\tau)\vert 0 \rangle,
\label{eq:gluecorr}
\end{equation}
where $\bar{\phi}^{(R)}_{\alpha}(t)$ denotes a vacuum-subtracted operator
$\bar{\phi}^{(R)}_{\alpha}(t) = \phi^{(R)}_{\alpha}(t) - \langle 0 \vert
 \phi^{(R)}_{\alpha}(t)\vert 0 \rangle$.  Note that $\langle 0 \vert
 \phi^{(R)}_{\alpha}(t)\vert 0 \rangle$ is independent of $t$.
The coefficients $v^{(R)}_\alpha$ were then determined by minimizing
the effective mass
\begin{equation}
\tilde m(t_D) = - \ln\left[
\frac{\sum_{\alpha\beta}
 v^{(R)}_\alpha v^{(R)}_\beta\ \tilde C_{\alpha\beta}(t_D)}
{\sum_{\alpha\beta}
 v^{(R)}_\alpha v^{(R)}_\beta\ \tilde C_{\alpha\beta}(0)}
\right],
\end{equation}
where the time separation for optimization was fixed in all cases to $t_D=1$.
Let ${\bf v}^{(R)}$ denote a column vector whose
elements are the optimal values of the coefficients $v^{(R)}_\alpha$.
Then requiring $d \tilde m(t_D) / d v^{(R)}_\alpha = 0$ for all
$\alpha$ yields an eigenvalue equation:
\begin{equation}
\tilde C(t_D)\ {\bf v}^{(R)} = e^{-\tilde m(t_D)}\ \tilde C(0)
\ {\bf v}^{(R)}. \label{eq:Variation}
\end{equation}
The eigenvector ${\bf v}_0^{(R)}$ corresponding to the largest eigenvalue
$e^{-\tilde m_0(t_D)}$ then yields the coefficients $v^{(R)}_{0\alpha}$
for the operator $\Phi^{(R)}_0(t)$ which best overlaps the lowest-lying
glueball $G$ in the channel of interest.  Operators which overlap excited
glueball states can also be constructed using the other eigenvectors of
Eq.~\ref{eq:Variation}.  In particular, the operator $\Phi^{(R)}_1(t)$
expected to best overlap the first-excited glueball state $G^\ast$ was
obtained from the eigenvector corresponding to the second largest eigenvalue
of Eq.~\ref{eq:Variation}.

The elements of the correlator matrix given in Eq.~\ref{eq:gluecorr}
were estimated using the Monte Carlo method.  Ten separate glueball
simulations were performed on DEC Alpha-workstations.
Configuration ensembles were generated using both Cabibbo-Marinari (CM)
pseudo-heatbath and $SU(2)$ sub-group over-relaxation (OR) methods.  Link
variables were updated in serial order on the lattice.  We define a {\em
compound sweep} as one CM updating sweep followed by three OR sweeps.  In
the glueball simulations, three compound sweeps were performed between
measurements, and the measurements were averaged into bins of 100 in order
to reduce data storage requirements (except for the $\beta=2.6$,
$\xi=3$ run in which 40 configurations were included in each bin).
In all ten simulations, 100 bins were obtained.  Our ensembles were tested
for residual autocorrelations during the analysis phase by over-binning by
factors of two and four; in all cases, the statistical error estimates
remained unchanged.

Values for the mean link parameter $u_s$ were determined self-consistently
as previously described. This tuning procedure required a minimal amount of
computational effort and provided thermalized configurations
for later computations.  The improved action simulation parameters used are
given in Table~\ref{tab:RunParams}.

For the data-fitting phase, the large $24\times 24$ correlator matrices in
each channel were reduced using the coefficients ${\bf v}_0^{(R)}$ and
${\bf v}_1^{(R)}$ to smaller $2\times 2$ matrices $C_{AB}(t)$ for $A,B = 0,1$:
\begin{equation}
C_{AB}(t)=\sum_\tau \langle 0\vert \bar{\Phi}_A^{(R)}(\tau\!+\!t)
\ \bar{\Phi}^{(R)}_B(\tau)\vert 0\rangle.
\end{equation}
The ground-state correlator $C_{00}(t)$ was fit
for $t=t_{\rm min},\dots,t_{\rm max}$ using a single exponential
\begin{equation}
  C_{00}(t) =  Z_{00}
        \left\{ e^{ -m_Gt} + e^{-m_G(T-t)} \right\},
\label{eq:GGcorrelator}
\end{equation}
where $T$ was the temporal extent of the periodic lattice, to obtain an
estimate of the mass $m_G$ (in terms of $a_t^{-1}$) of the lowest-lying
glueball in each channel. To determine the mass $m_{G^\ast}$ of the
first-excited glueball and another estimate of $m_G$, the $2\times 2$
correlator was also fit for $t=t_{\rm min},\dots,t_{\rm max}$ using the form
\begin{equation}
  C_{AB}(t) = \sum_{p=G,G^\ast}   Z_{Ap} Z_{Bp}
        \left\{ e^{ -m_pt} + e^{-m_p(T-t)} \right\}.
\label{eq:2by2correlator}
\end{equation}
Various fit regions $t_{\rm min}$ to $t_{\rm max}$ were used in order to
check for consistency in the extracted values for the masses.
Best fit values were obtained using the correlated $\chi^2$ method.
Error estimates were calculated using a $1024-$point bootstrap procedure;
in all cases, error estimates were very close to being symmetric about
the central best-fit values and were thus averaged to simplify presentation.

\section{Setting the scale using the static potential}
\label{sec:r0as}

In order to convert the glueball masses as measured in our
simulations into physical units, we must set the scale by determining
the lattice spacing $a_t$ for each $\beta$ and $\xi$ we consider.
To do this, we must first choose one physical quantity to use
as a reference.  This reference quantity must then be measured
on the lattice in terms of $a_t$.  The experimentally-known
value for the reference quantity is then used to extract the
lattice spacing.  A quantity which can be easily and accurately
determined both experimentally and in numerical simulations is
an ideal choice for such a reference.  The mass of a low-lying
particle is typically used for setting the scale.  In our case,
however, there are no unambiguous experimental determinations of the glueball 
masses, so instead, we must look for another purely gluonic quantity.

The hadronic scale parameter $r_0$ defined in terms of the force between
static quarks by $[r^2 dV(\vec{r})/dr]_{r=r_0}=1.65$, where $V(\vec{r})$
is the static-quark potential, is an attractive possibility.
It can be measured very accurately on the lattice.  The advantages
in using $r_0$ to set the scale have been enumerated in
Ref.~\cite{sommer}.  From phenomenological potential models, one
finds $r_0\approx 0.5$ fm.  A disadvantage in using $r_0$ is that its
physical value must be deduced indirectly from experiment, and
there is some ambiguity in doing this, as will be discussed below.
However, in the absence of a better gluonic reference, we have
chosen $r_0$ to set the scale.  In this section, we outline
the determination of $r_0$ in terms of $a_s$.

In order to determine $r_0$ in terms of the lattice spacing, we need
accurate measurements of the static-quark potential.  We extracted $V(\vec{r})$
for various spatial separations $\vec{r}$, both on and off the axes of the
lattice, from the expectation values of Wilson loops $W(\vec{r},t)$
in the standard manner:
\begin{equation}
 W(\vec{r},t)=Z(\vec{r})\ \exp[-tV(\vec{r})]
 + {\rm excited\ state\ contributions}.
\end{equation}
In the Monte Carlo evaluation of the Wilson loops, measurements
were taken after every four compound sweeps (as defined in
Sec.~\ref{sec:simulation}).   The measurements of the Wilson
loops were done independently of the glueball mass studies
using separate ensembles of configurations.  To minimize contamination
from excited states, the Wilson loops were constructed from iteratively
smeared spatial links.  The single-link smearing method described previously
was used.  A given smearing scheme is specified not only by the
parameter $\lambda_s$, but also by the total number of smearing iterations,
denoted by $n_\lambda$.  Two different choices of the smearing parameter
were used in all cases: one smearing was chosen to work well for small
$r=\vert\vec{r}\vert$, the other to work well for large $r$.
Separate measurements for each smearing were taken; cross correlations
were not determined.  The statistical noise in the evaluation of $W(\vec{r},t)$
was reduced dramatically, especially for large temporal separations,
by constructing the Wilson loops, whenever possible, from thermally-averaged
temporal links \cite{thermal}.  The thermal averaging was accomplished using
the Cabibbo-Marinari pseudo-heatbath method (40 updates).  Other relevant run
parameters are given in Table~\ref{potrun}.

\begin{figure}
\begin{center}
\leavevmode
\epsfxsize=5in\epsfbox[80 140 530 760]{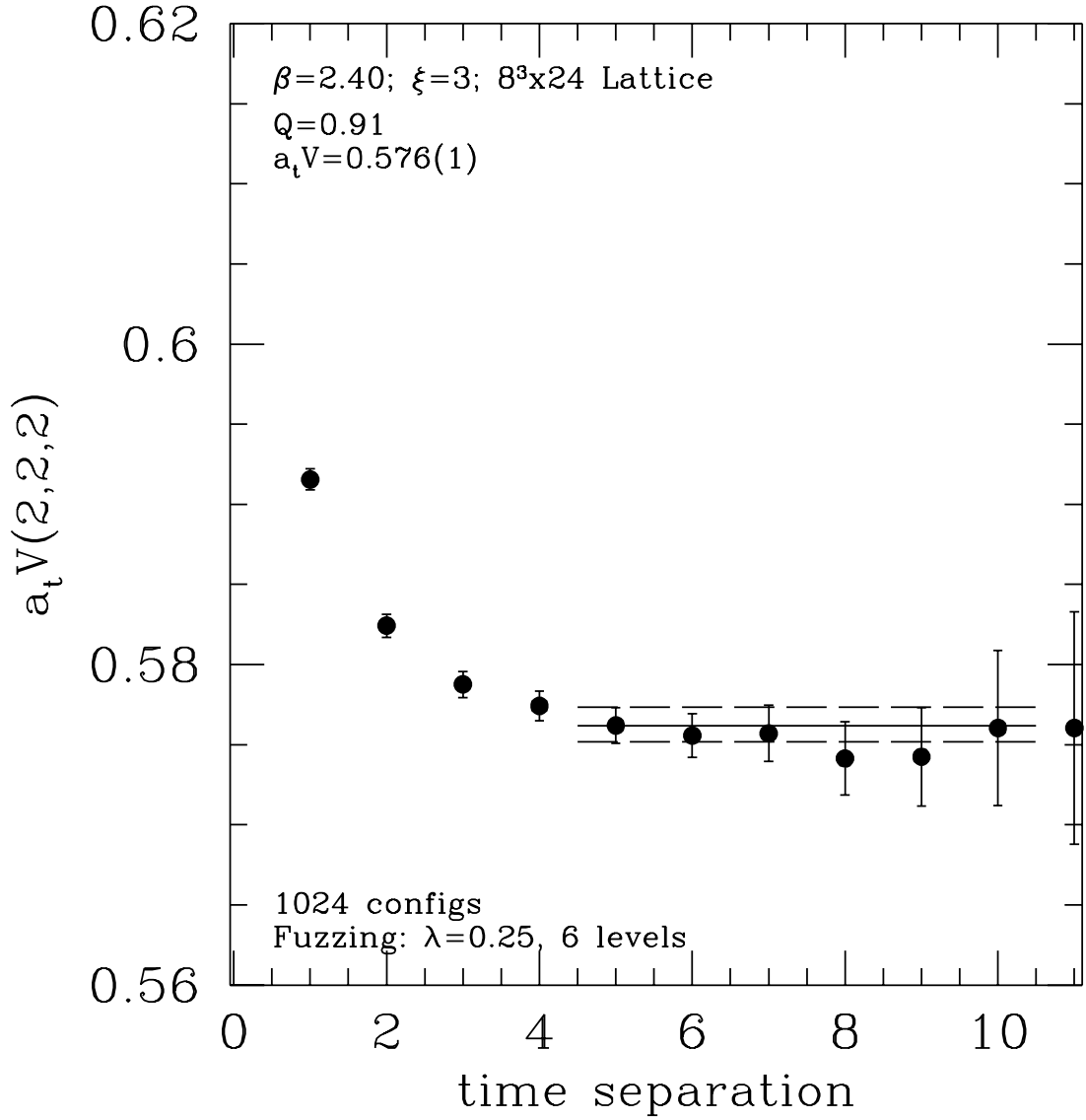}
\end{center}
\caption[Veff]{
 Effective mass plot showing the results of a single-exponential
 fit to the Wilson loop for $V(\vec{r})$ with $\vec{r}/a_s=(2,2,2)$,
 $\beta=2.4$, and $\xi=3$.  The $t_{\rm min}-t_{\rm max}$ region of the fit
 is also indicated.
\label{Veffmass}}
\end{figure}

The values of the potential $V(\vec{r})$ were extracted from the Wilson
loop measurements by fitting $W(\vec{r},t)$ to the exponential
form $Z(\vec{r})\exp[-tV(\vec{r})]$ in the range $t=t_{\rm min},\dots,
t_{\rm max}$,
for each $\vec{r}$.  The plateau region from $t_{\rm min}$ to $t_{\rm max}$ was
chosen separately for each $\vec{r}$ in order to minimize the uncertainty
in the extracted values for $V(\vec{r})$ while maintaining a good quality
$Q$ of fit.  Best fit values were determined using the standard
$\chi$-square test, taking into account temporal correlations
among the $W(\vec{r},t)$.  The covariance matrix in $\chi^2$ was determined
using the jack-knife procedure, and estimates for the uncertainties
in the extracted values for $V(\vec{r})$ were computed using the bootstrap
method.  Binning of the data was done as a crude check that our
measurements were statistically independent.  The results of a typical
fit are shown in Fig.~\ref{Veffmass}, which is an effective mass
plot for $V(\vec{r})$ for $\vec{r}/a_s=(2,2,2)$.  The effective mass
for $V(\vec{r})$ is a function of $t$, defined as
$\ln[W(\vec{r},t)/W(\vec{r},t+a_t)]$, which tends to the true mass
as $t$ becomes large.

Once a suitable plateau region in the effective mass was established
for each $V(\vec{r})$, the hadronic scale $r_0/a_s$ could be determined.
We found that the on-axis potential $V(\vec{r})$ for the range
of $\vec{r}$ values studied here using coarse lattices fit a
Coulomb plus linear form $V(\vec{r})=e_c/r+\sigma r+V_0$ very well
(with qualities of fit ranging from $Q=0.25$ to $Q=0.99$).
We, therefore, used this form to interpolate $V(\vec{r})$ and the
force between static quarks.  Simultaneous fits of the Wilson
loops for the on-axis potential to the form
$Z(r)\ \exp[-t(e_c/r+\sigma r+V_0)]$ were done, taking into
account all correlations among the $W(r,t)$ for both different
$t$ and $r$.  Different regions in $t$ were used for
different $r$ values; the plateau regions determined previously
were used.  Only the on-axis potentials were used; this
prevented the covariance matrix in the $\chi^2$ to be
minimized from getting too large.  This covariance matrix
was evaluated using the jack-knife method; uncertainties in
the fit parameters $e_c$, $\sigma$, $V_0$, and $Z(r)$ were
obtained using the bootstrap method.  Once we had an ensemble
of bootstrap estimates for these fit parameters, the ratio $r_0/a_s$ and
its bootstrap uncertainty were then determined using
\begin{equation}
r_0/a_s = \sqrt{(1.65 + e_c)/\sigma a_s^2}.
\end{equation}
Note that to compute $r_0/a_s$, we need the ratio $a_s/a_t$ since our
fits yielded estimates of $a_tV(\vec{r})$ only.  We used the input value
$\xi$ since we know that its renormalization is small.  Results for
$r_0/a_s$ are given in Table~\ref{r0values}.

\begin{figure}
\begin{center}
\leavevmode
\epsfxsize=5in\epsfbox[80 160 530 760]{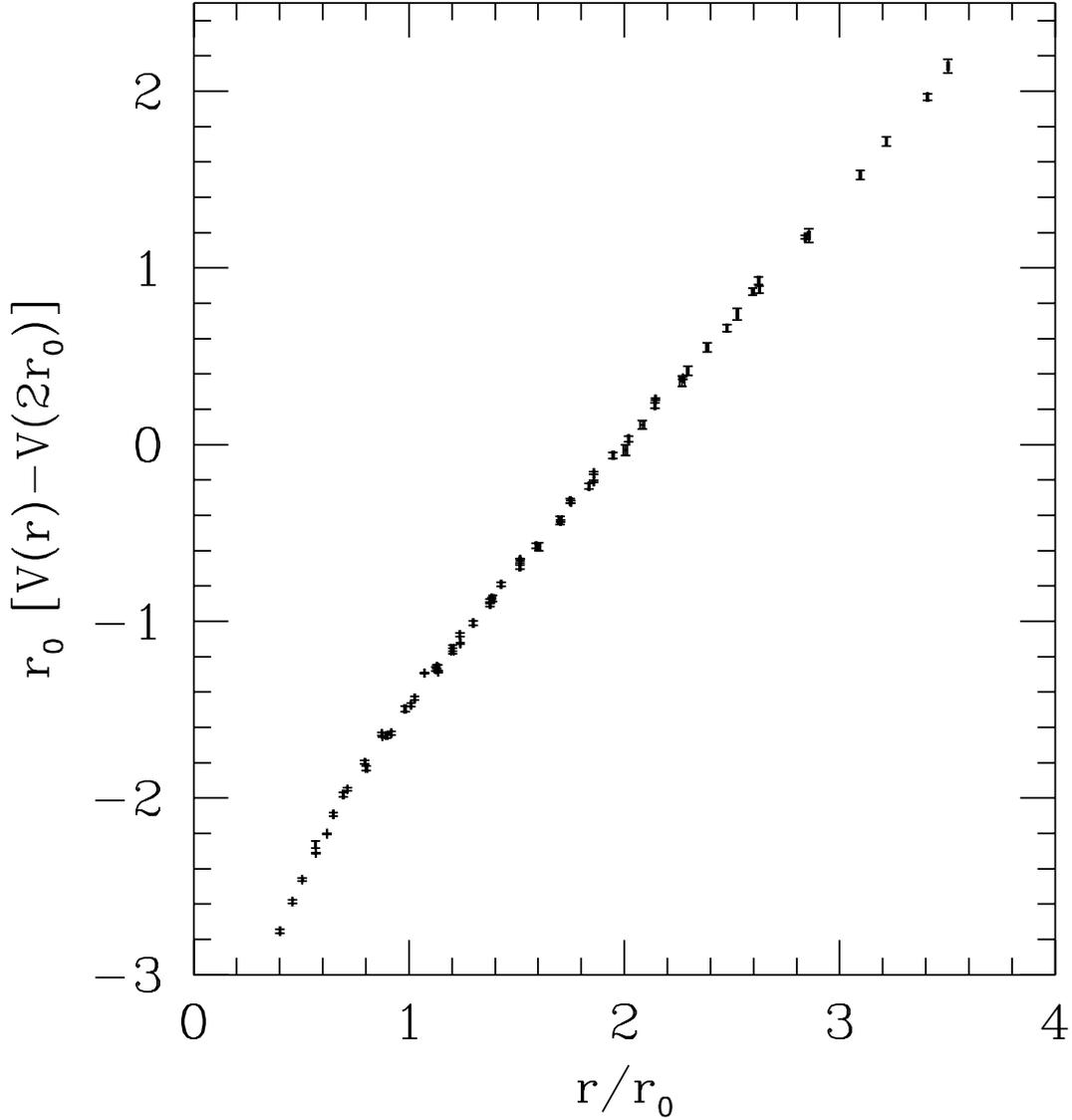}
\end{center}
\caption[VR]{
 The static-quark potential $V(\vec{r})$ expressed in terms of the hadronic
 scale $r_0$.  This plot includes measurements from the
 $\beta=2.2$, $2.4$, and $2.6$ simulations for $\xi=3$, and
 the $\beta=2.2$ and $2.4$ simulations for $\xi=5$.  Lattice
 spacing errors are seen to be small.
\label{potential}}
\end{figure}

Using the results in Table~\ref{r0values}, we can now express
all energies measured in simulations in terms of $r_0$.
For example, in Fig.~\ref{potential}, we show the potential, including
off-axis inter-quark separations, expressed in terms of $r_0$.
Lattice spacing errors are seen to be small.

\section{Results and Discussion}
\label{sec:resdiscuss}

\subsection{Glueball mass measurements}
\label{sec:results}

To allow clear resolution of the scaling properties of the low-lying
glueball masses in the improved action, two sets of simulations
were performed at two different anisotropies: six lattice spacings
for an aspect ratio $\xi=3$ and four spacings for $\xi=5$ were
studied.  The input parameters used in these simulations are given
in Table~\ref{tab:RunParams}.

\begin{figure}
\begin{center}
\leavevmode
\epsfxsize=5.5 true in\epsfbox{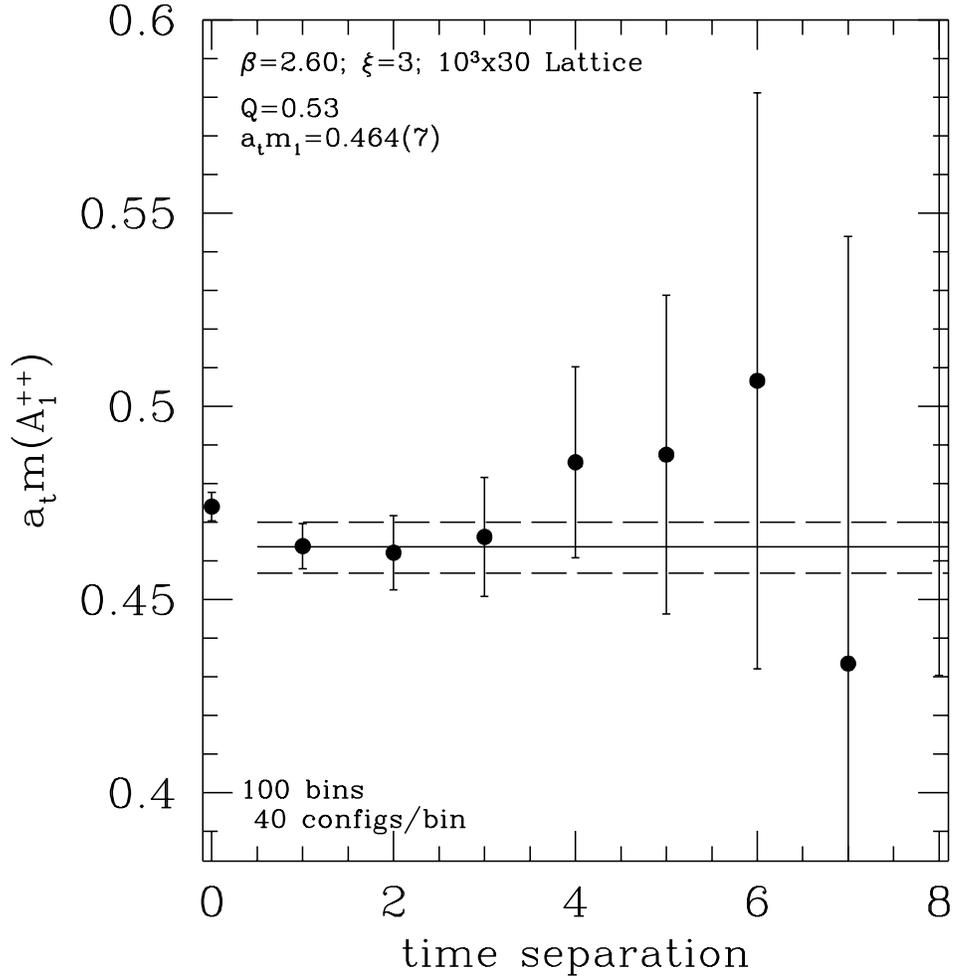}
\end{center}
\caption[figemone]{
 Effective mass plot showing the results of a single-exponential
 fit to the glueball correlation function for the $A_1^{++}$ channel
 for $\beta=2.6$ and $\xi=3$.  The $t_{\rm min}-t_{\rm max}$ region of the
 fit is also indicated.
\label{fig:B26-R3-effmass-a1pp}}
\end{figure}

\begin{figure}
\begin{center}
\leavevmode
\epsfxsize=5.5 true in\epsfbox{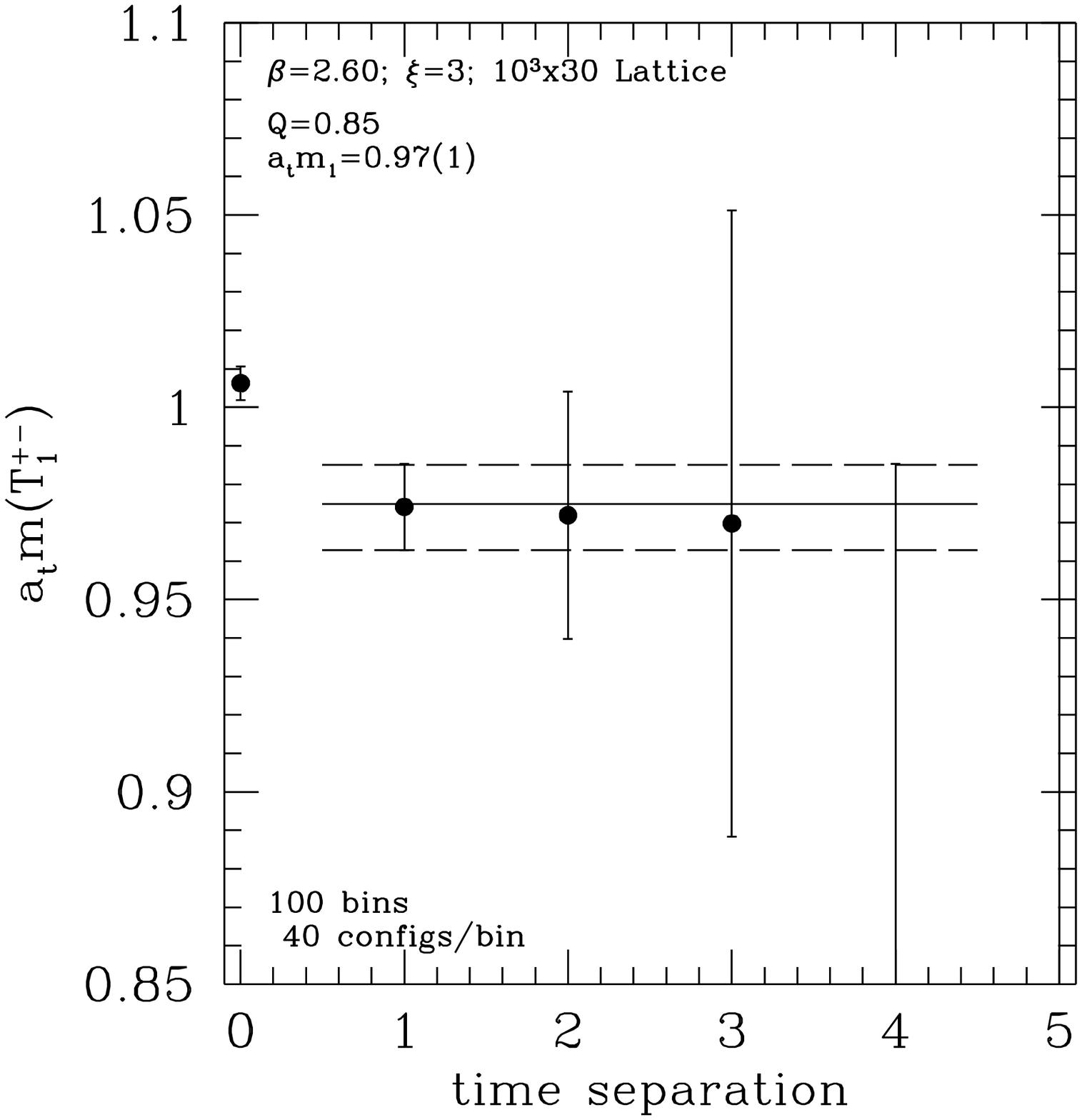}
\end{center}
\caption[figemone]{
 Effective mass plot showing the results of a single-exponential
 fit to the glueball correlation function for the $T_1^{+-}$ channel
 for $\beta=2.6$ and $\xi=3$.
\label{fig:B26-R3-effmass-t1pm}}
\end{figure}

\begin{figure}
\begin{center}
\leavevmode
\epsfxsize=5.5 true in\epsfbox{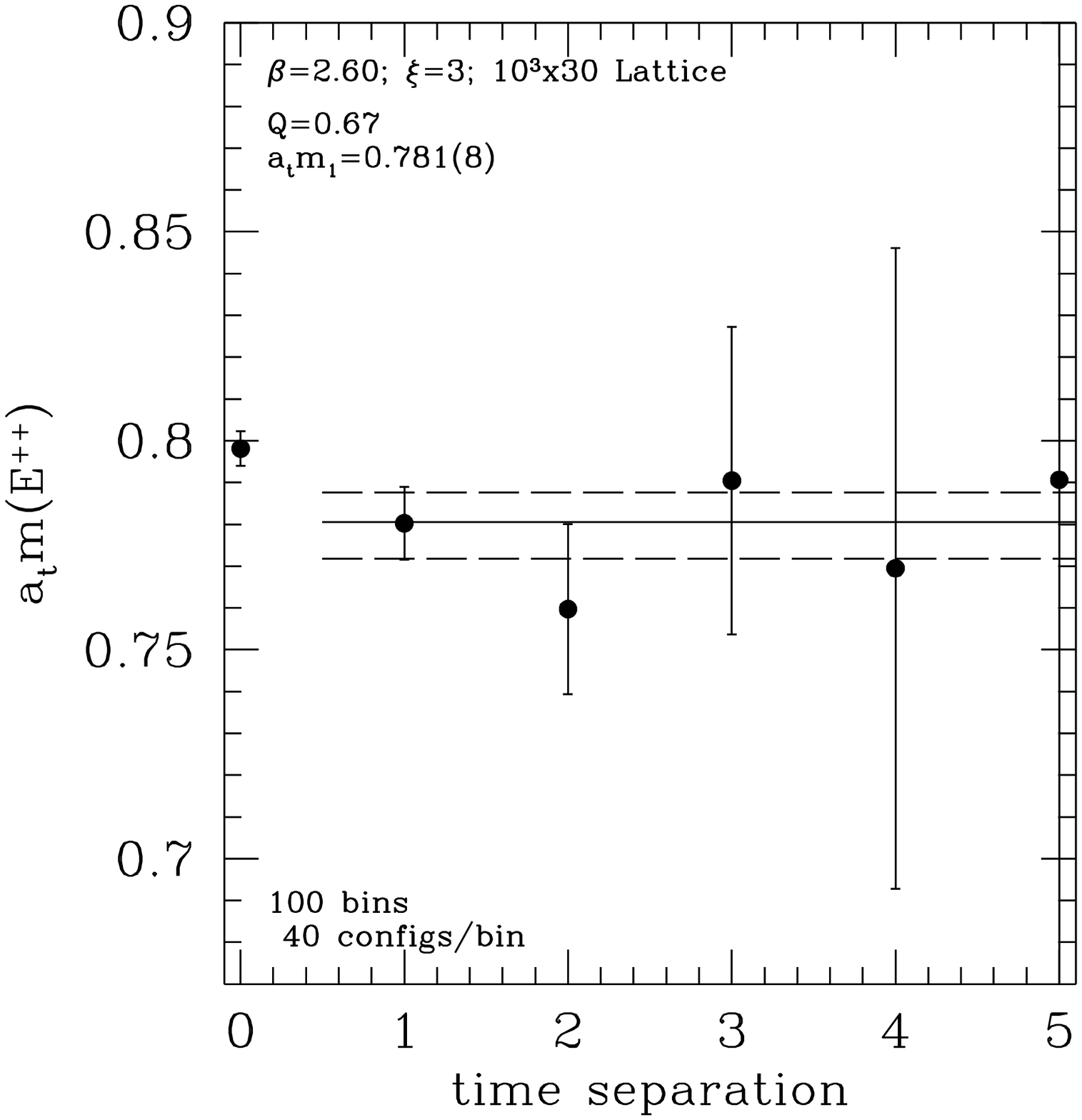}
\end{center}
\caption[figemone]{
 Effective mass plot showing the results of a single-exponential
 fit to the glueball correlation function for the $E^{++}$ channel
 for $\beta=2.6$ and $\xi=3$.
\label{fig:B26-R3-effmass-e-pp}}
\end{figure}

\begin{figure}
\begin{center}
\leavevmode
\epsfxsize=5.5 true in\epsfbox{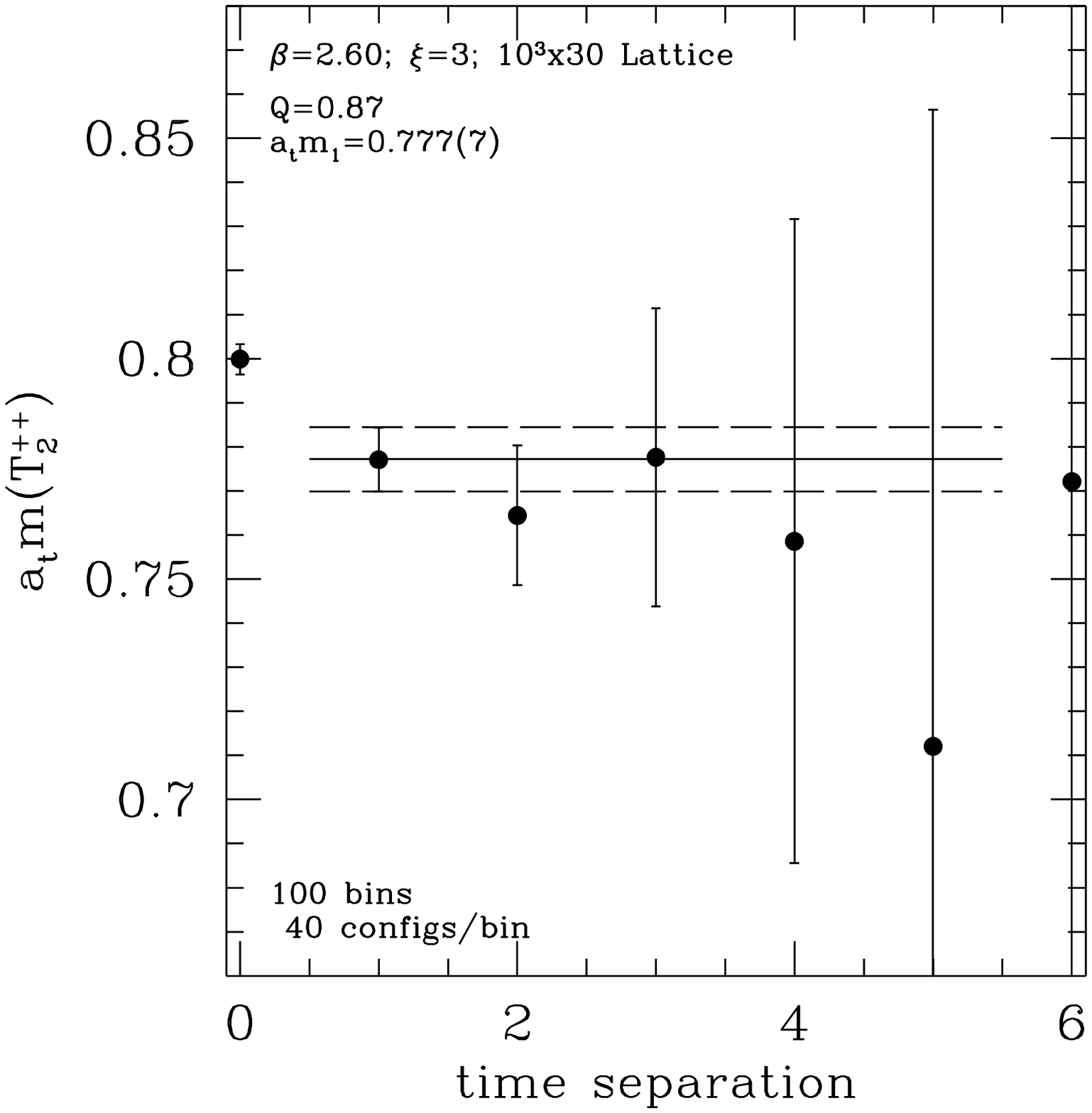}
\end{center}
\caption[figemone]{
 Effective mass plot showing the results of a single-exponential
 fit to the glueball correlation function for the $T_2^{++}$ channel
 for $\beta=2.6$ and $\xi=3$.
\label{fig:B26-R3-effmass-t2pp}}
\end{figure}

\begin{figure}
\begin{center}
\leavevmode
\epsfxsize=400pt\epsfbox{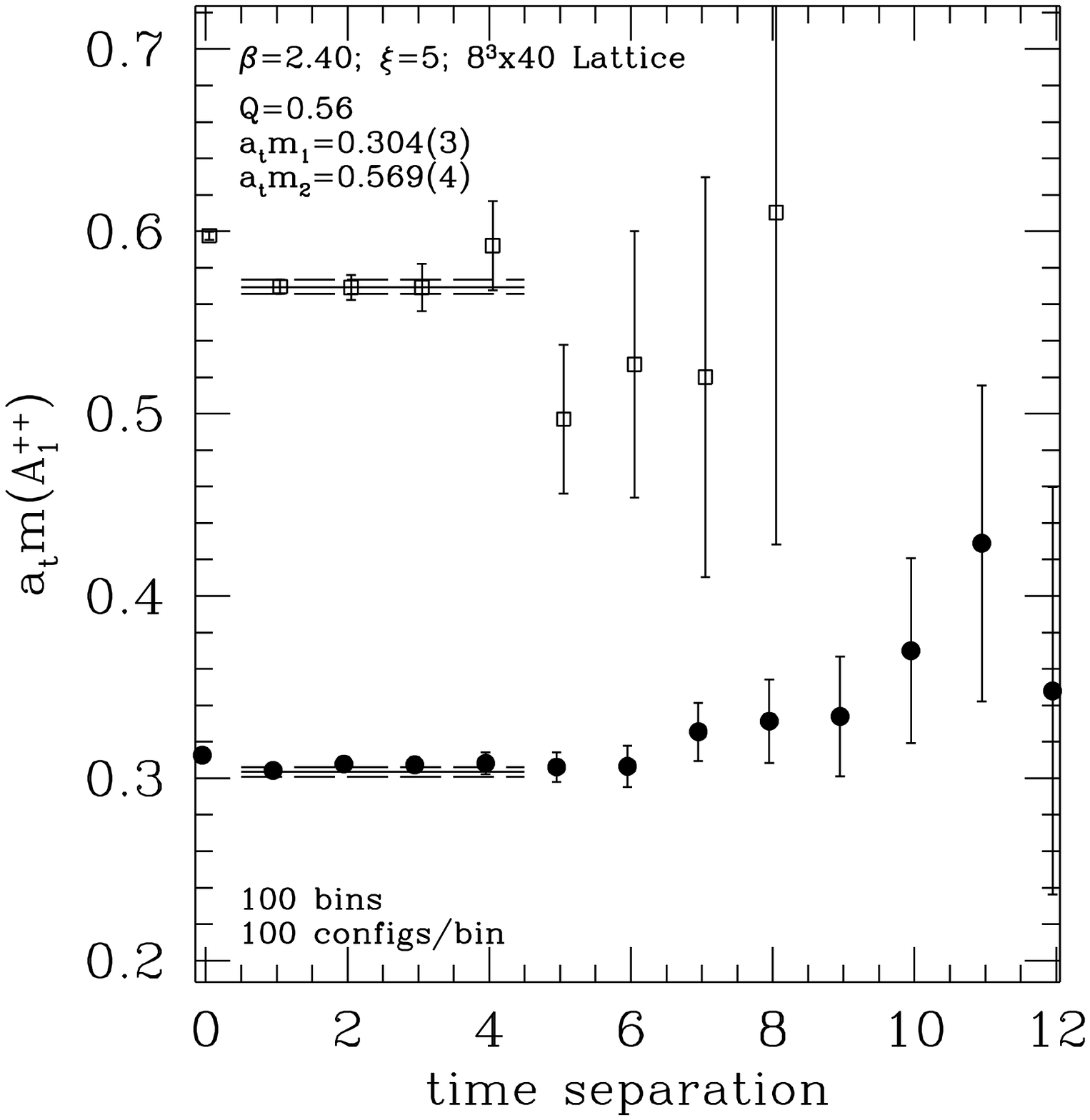}
\end{center}
\caption[figemone]{
 Effective mass plot showing the results of a two-exponential
 fit to the $2\times 2$ matrix of glueball correlation functions for
 the $A_1^{++}$ channel for $\beta=2.4$ and $\xi=5$.  The
 $t_{\rm min}-t_{\rm max}$ region of the fit is also indicated.
\label{fig:B24-R5-effmass-a1pp}}
\end{figure}

\begin{figure}
\begin{center}
\leavevmode
\epsfxsize=400pt\epsfbox{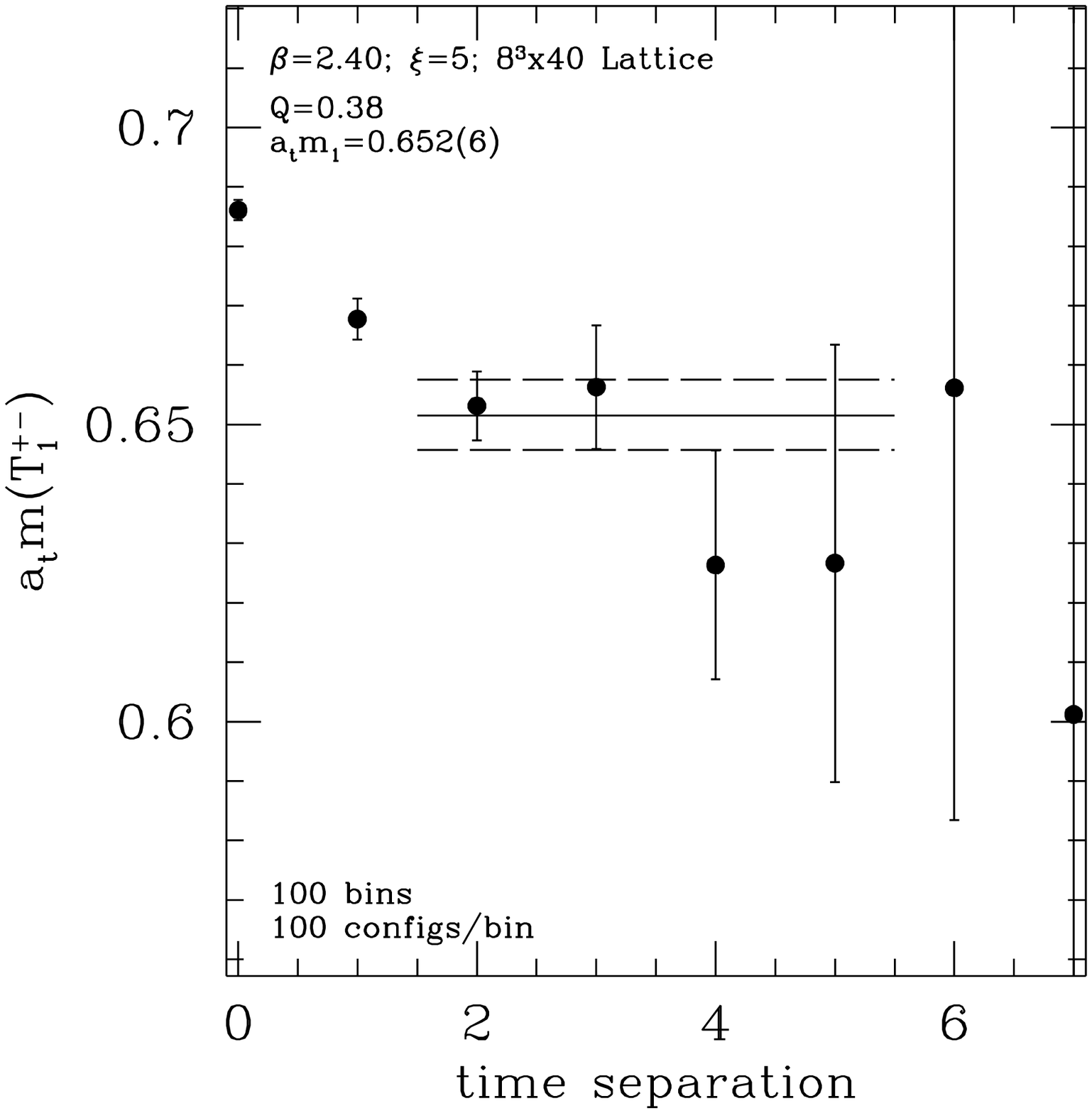}
\end{center}
\caption[figemone]{
 Effective mass plot showing the results of a single-exponential
 fit to the glueball correlation function for the $T_1^{+-}$ channel
 for $\beta=2.4$ and $\xi=5$.
\label{fig:B24-R5-effmass-t1pm}}
\end{figure}

\begin{figure}
\begin{center}
\leavevmode
\epsfxsize=400pt\epsfbox{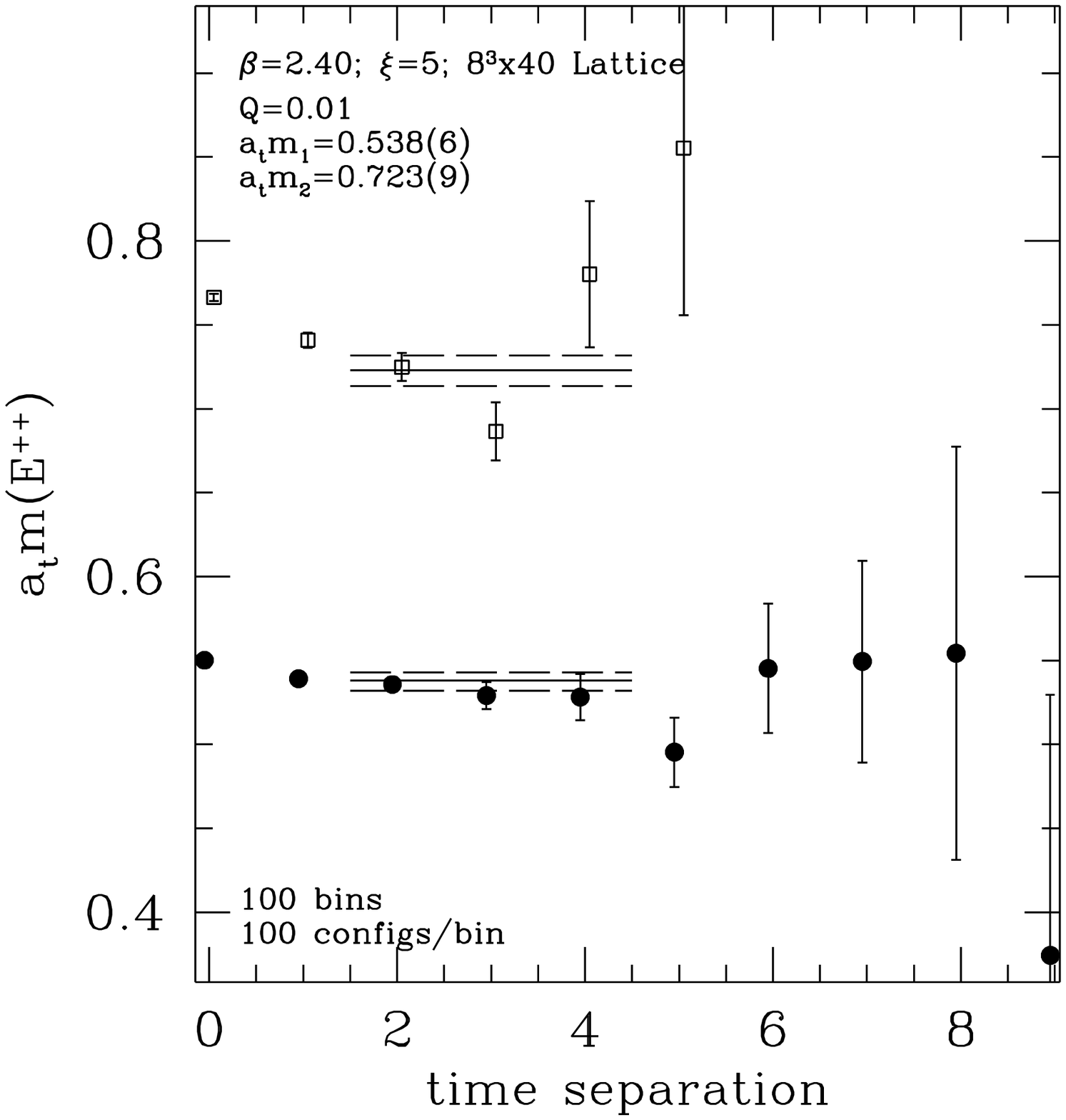}
\end{center}
\caption[figemone]{
 Effective mass plot showing the results of a two-exponential
 fit to the $2\times 2$ matrix of glueball correlation functions for
 the $E^{++}$ channel for $\beta=2.4$ and $\xi=5$.  The
 $t_{\rm min}-t_{\rm max}$ region of the fit is also indicated.
\label{fig:B24-R5-effmass-e-pp}}
\end{figure}

\begin{figure}
\begin{center}
\leavevmode
\epsfxsize=400pt\epsfbox{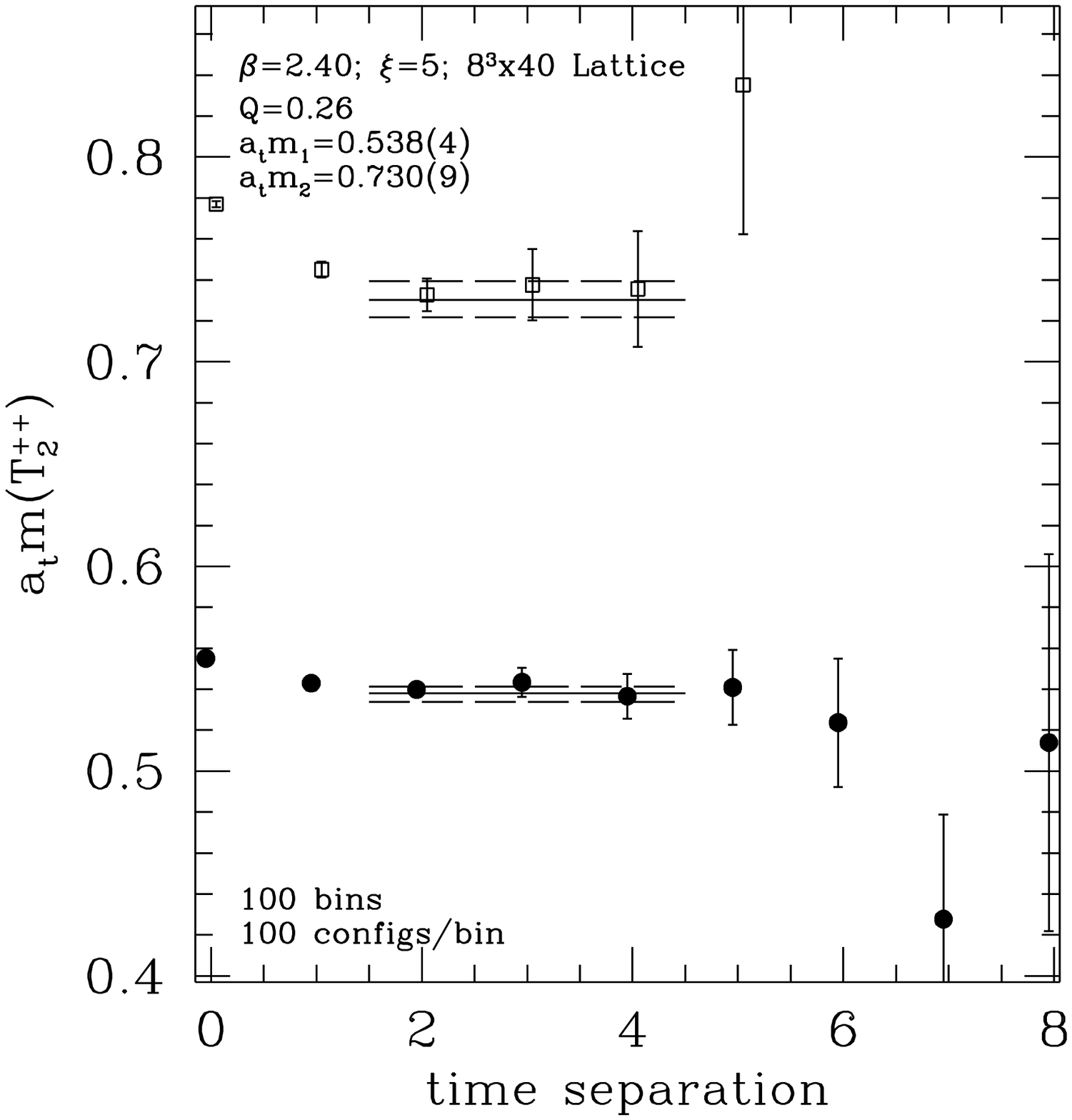}
\end{center}
\caption[figemone]{
 Effective mass plot showing the results of a two-exponential
 fit to the $2\times 2$ matrix of glueball correlation functions for
 the $T_2^{++}$ channel for $\beta=2.4$ and $\xi=5$.  The
 $t_{\rm min}-t_{\rm max}$ region of the fit is also indicated.
\label{fig:B24-R5-effmass-t2pp}}
\end{figure}

The results of fitting the variationally-optimized correlators $C(t)$ to
the functions given in Eqs.~\ref{eq:GGcorrelator} and \ref{eq:2by2correlator}
are summarized in Tables~\ref{tab:emone}-\ref{tab:emten}.  Effective mass
plots for the two smallest-$a_s$ simulations are presented in
Figs.~\ref{fig:B26-R3-effmass-a1pp}$-$\ref{fig:B26-R3-effmass-t2pp} and
Figs.~\ref{fig:B24-R5-effmass-a1pp}$-$\ref{fig:B24-R5-effmass-t2pp}
for $\xi=3$ and $\xi=5$, respectively.  For each
channel in each of the ten simulations, it was possible to find a fit region
$t_{\rm min}-t_{\rm max}$ in which the correlation function was well
described by its asymptotic form as indicated by the quality of fit.
In other words, convincing plateaux were observed in all effective masses.
The most impressive plateau, observed in the $A_1^{++}$ channel for
$\beta=2.4$ and $\xi=5$, spanned ten time-slices.  In most cases, the onset
of the plateau occurred when the source and sink operators were separated by
only one time-step.  The overlaps with the lowest-lying states were also
found to be extremely good, better than $90\%$ in most cases and often
consistent with unity.  This clearly demonstrates the effectiveness of
the link-smearing and variational techniques in diminishing excited-state
contamination.  Fits using $t_{\rm min}= 2$ or $0$ were also done to check
for agreement with the $t_{\rm min}=1$ results.  At time separations for
which the ground state could be reliably observed, the off-diagonal elements
of the reduced correlation matrices $C_{AB}(t)$ were found to be consistent
with zero within statistical uncertainty.  This suggests that the
link-smearing, variational method also gives an excellent construction of
the first-excited state in each channel.

Our best estimates for the glueball masses in terms of $a_t^{-1}$ are
indicated in boldface in each of the Tables~\ref{tab:emone}-\ref{tab:emten}.
These estimates are summarized in Table~\ref{tab:final}.  Masses for
the first-excited states are also indicated in the $N_{exp}=2$ fits
listed in the Tables~\ref{tab:emone}-\ref{tab:emten}.

\subsection{Finite volume effects}

In this work, we were concerned with the magnitude of discretization errors
in the glueball mass determinations from coarse lattice simulations
using an improved action.  In order to evaluate these errors, we had to
eliminate uncertainties from all other sources.  The increased efficiency
of simulations on coarse, anisotropic lattices allowed us to reduce
statistical errors to the acceptable level of about $1\%$.  The only
remaining source of uncertainty we had to address was finite volume.
The masses of particles confined in a small box with periodic boundary
conditions can differ appreciably from their infinite volume values;
finite volume effects can also induce a splitting in the masses of the
$E$ and $T_2$ tensor polarizations.  Finite volume effects on the
scalar glueball mass have been analyzed before\cite{Schierholz89}, but
the effects on the tensor and the pseudovector are less well known.

In order to ascertain the effects on our glueball masses of simulating
in a finite volume, four extra simulations were performed for $\beta=2.4$,
$\xi=3$ using lattices of spatial extent $L_s/a_s = 6, 5, 4,$ and $3$.
The temporal extent was held fixed at 24 grid points.  For each of these
volumes, the mean-field renormalization parameter $u_s$ was recalculated.
The $3^3$ lattice was the only simulation that required any change in this
parameter, and in this case, the effect was small; $u_s$ increased by only
$0.3\%$.  The results from the $L_s/a_s=8,6,5,$ and $4$ runs for the glueball
masses in terms of $a_t^{-1}$ are given in Table~\ref{tab:FinVolVals}.  Note
that the results from the $6^3$ lattice differ very little from those from
the $8^3$ lattice, suggesting that our lattice volumes are sufficiently
large to ensure that finite volume errors are negligible.  For the $3^3$
lattice, no plateaux in the effective masses for the $A_1^{++}$, $E^{++}$ and
$T_2^{++}$ channels were observed; the mass in the $T_1^{+-}$ channel
was found to be $1.44(1)$.   The operators used in these runs were the
same as those constructed for the large volume runs and thus, were not
optimized to give large overlaps with the light torelon states present in
small volumes.  It is likely that this effect was responsible for the poor
overlap of our operators with the scalar and tensor states on the
$3^3$ lattice.

The properties, such as the mass, of a glueball confined in a small box with
periodic boundary conditions differ from those in an infinite volume.  The
modification of the mass of a particle due to finite volume effects has been
estimated in Ref.~\cite{Luscher86}:
\begin{equation}
a_t m_G(z)  = a_t m_G(\infty)  \left[ 1 - \lambda_G
\exp(-\sqrt{3}z/2)/z  \right],
\label{eq:LuscherFinVol}
\end{equation}
where $z$ is the dimensionless length scale $z = m_{A_1^{++}}L_s$,
$m_{A_1^{++}}$ is the infinite-volume mass of the scalar glueball, and
$\lambda_G$ is related to the strength of an effective triple scalar glueball
interaction vertex.  The mass shift given in Eq.~\ref{eq:LuscherFinVol} is
valid for sufficiently large $z$ and arises from the exchange of scalar
glueballs across the periodic boundaries of the lattice.
Finite volume errors in our glueball masses measured on an $8^3$
lattice at $\beta=2.4$, $\xi=3$ (where $L_s \approx 2$ fm, similar to the
volumes used in the other nine simulations) can be estimated by fitting
the form given in Eq.~\ref{eq:LuscherFinVol} to the masses in
Table~\ref{tab:FinVolVals}.  Let $\omega=a_t m_{A_1^{++}}$ and
$\rho=\xi L_s/a_s$, then $z=\rho \omega$.  The $A_1^{++}$ fit was done
first using the function $a_t m(z)=\omega-\lambda_{A_1^{++}}\exp(-\sqrt{3}\rho
\omega/2)/\rho$, where $\omega$ and $\lambda_{A_1^{++}}$ were the
fitting parameters.  The best fit value for $\omega$ was then used
in the fits to the results for the other irreps; to simplify matters,
the uncertainty in $\omega$ was neglected in these fits.  The $T_1^{+-}$
fit also included the energy estimate extracted from the $L_s/a_s=3$
simulation.  The results of these fits are summarized in
Table~\ref{tab:FinVolFits}; the estimates of the finite volume errors are
listed in the final column of this table and are given by $m_G(8\omega\xi)/
m_G(\infty)-1$ using Eq.~\ref{eq:LuscherFinVol}.  In all cases, these errors
were insignificant compared to statistical errors;  this means, for example,
that any differences between the large-volume masses in the $T_2$ and $E$
channels must be due purely to discretization errors.  It is interesting to
note that our estimate of $\lambda_{A_1^{++}}$ agrees well with the value
$190\pm 70$ found in Ref.~\cite{Schierholz89}.

\subsection{Continuum limit extrapolations}

\begin{figure}
\begin{center}
\leavevmode
\epsfxsize=400pt\epsfbox{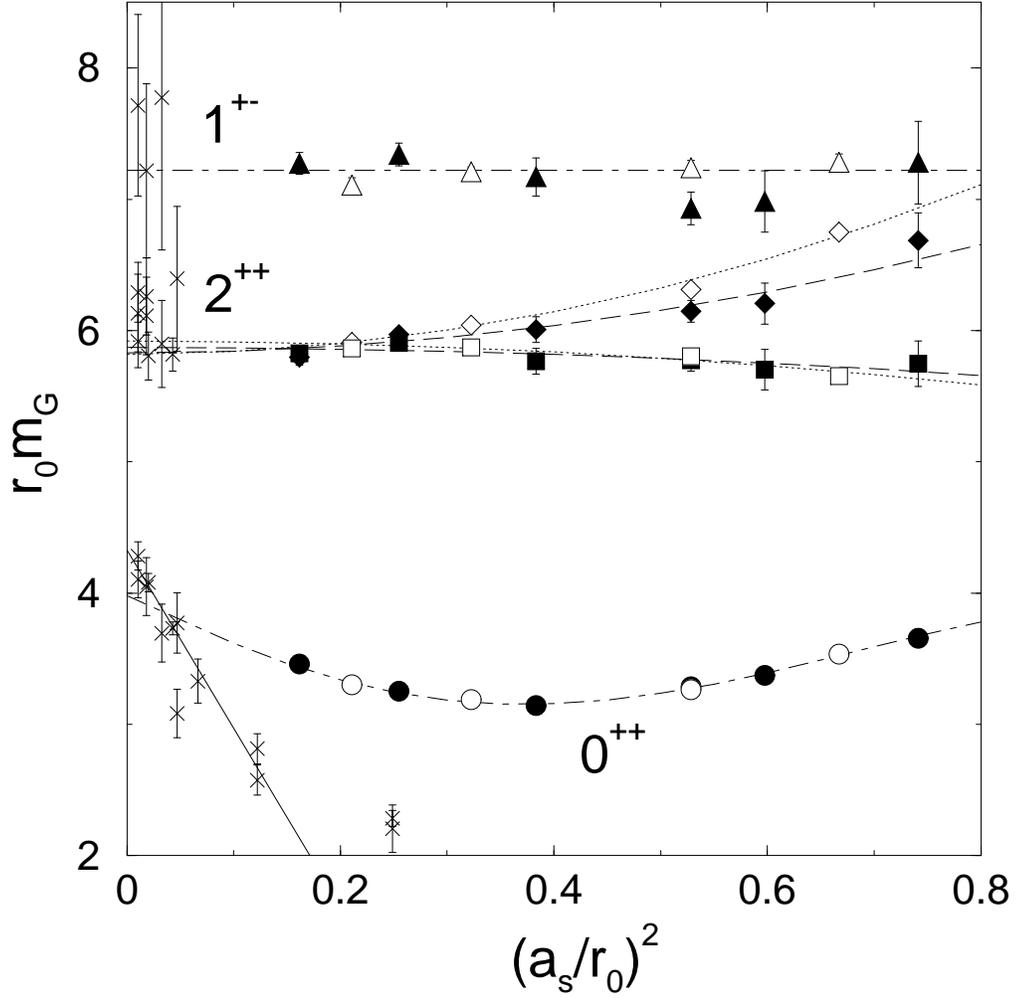}
\end{center}
\caption[figscale]{
Glueball mass estimates in terms of $r_0$ against the lattice
spacing $(a_s/r_0)^2$.  Results from the $\xi=5$ simulations for the
lattice irreps $A_1^{++}$, $E^{++}$, $T_2^{++}$ and $T_1^{+-}$
are labeled $\circ, \Box, \Diamond$, and $\triangle$, respectively.
The corresponding solid symbols indicate the results from the
$\xi=3$ simulations.  Data from Wilson action simulations taken from
Refs.~\protect{\cite{Chen,CMMT,UKQCD,Forcrand}} are shown using crosses.
The dashed, dotted, and dash-dotted curves indicate extrapolations to
the continuum limit obtained by fitting to the $\xi=3$ data, the $\xi=5$
data, and all data, respectively.  The solid line indicates the
extrapolation of the Wilson action data to the continuum limit.
\label{fig:Scaling}}
\end{figure}

\begin{figure}
\begin{center}
\leavevmode
\epsfxsize=400pt\epsfbox{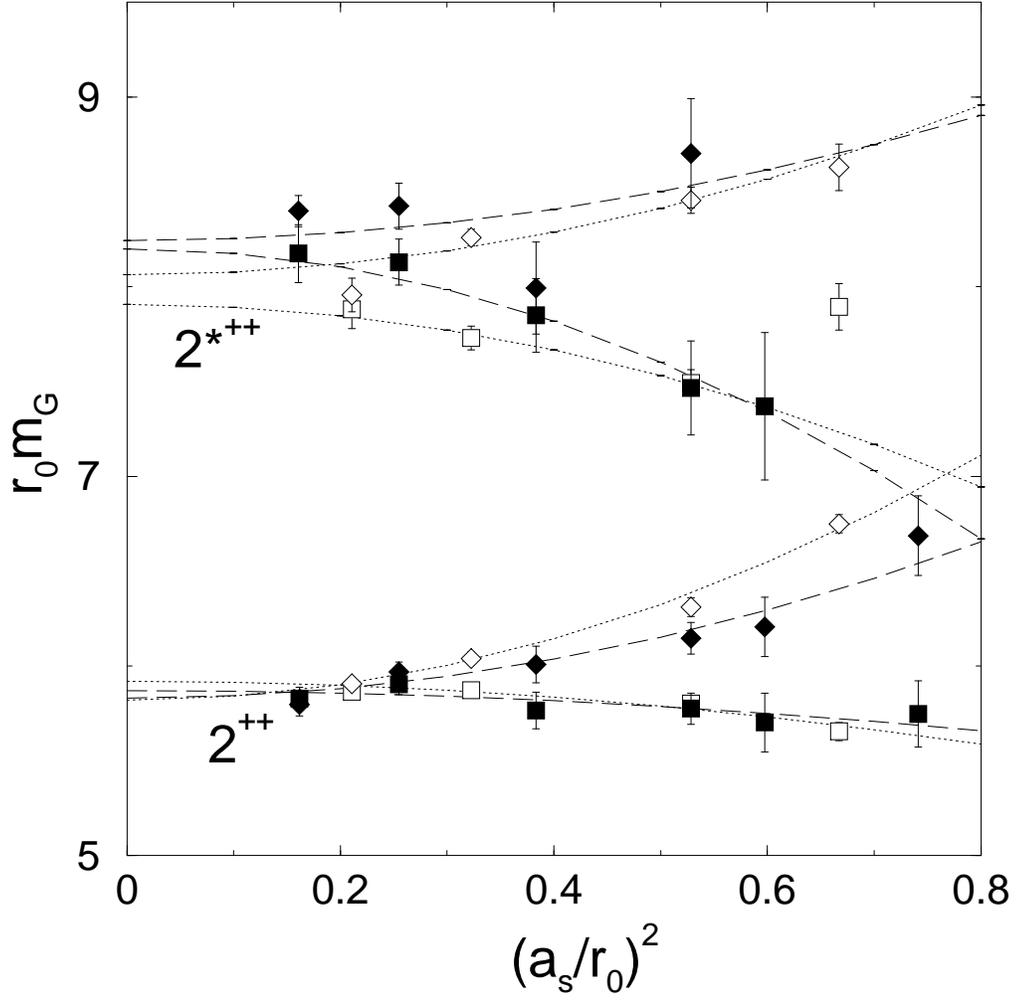}
\end{center}
\caption[figtensor]{
Estimates of the masses of the tensor glueball and its first-excited state
in terms of $r_0$ against the lattice spacing $(a_s/r_0)^2$.  Results
from the $\xi=5$ simulations for the $E^{++}$ and $T_2^{++}$ irreps are
labeled by $\Box$ and $\Diamond$, respectively.  The corresponding
solid symbols show the results from the $\xi=3$ simulations.
The dashed and dotted curves indicate extrapolations to the continuum
limit obtained by fitting to the $\xi=3$ and the $\xi=5$, respectively
(see Tables~\protect{\ref{tab:Extrap-3}} and \protect{\ref{tab:Extrap-5}}).
\label{fig:tensor}}
\end{figure}

\begin{figure}
\begin{center}
\leavevmode
\epsfxsize=400pt\epsfbox{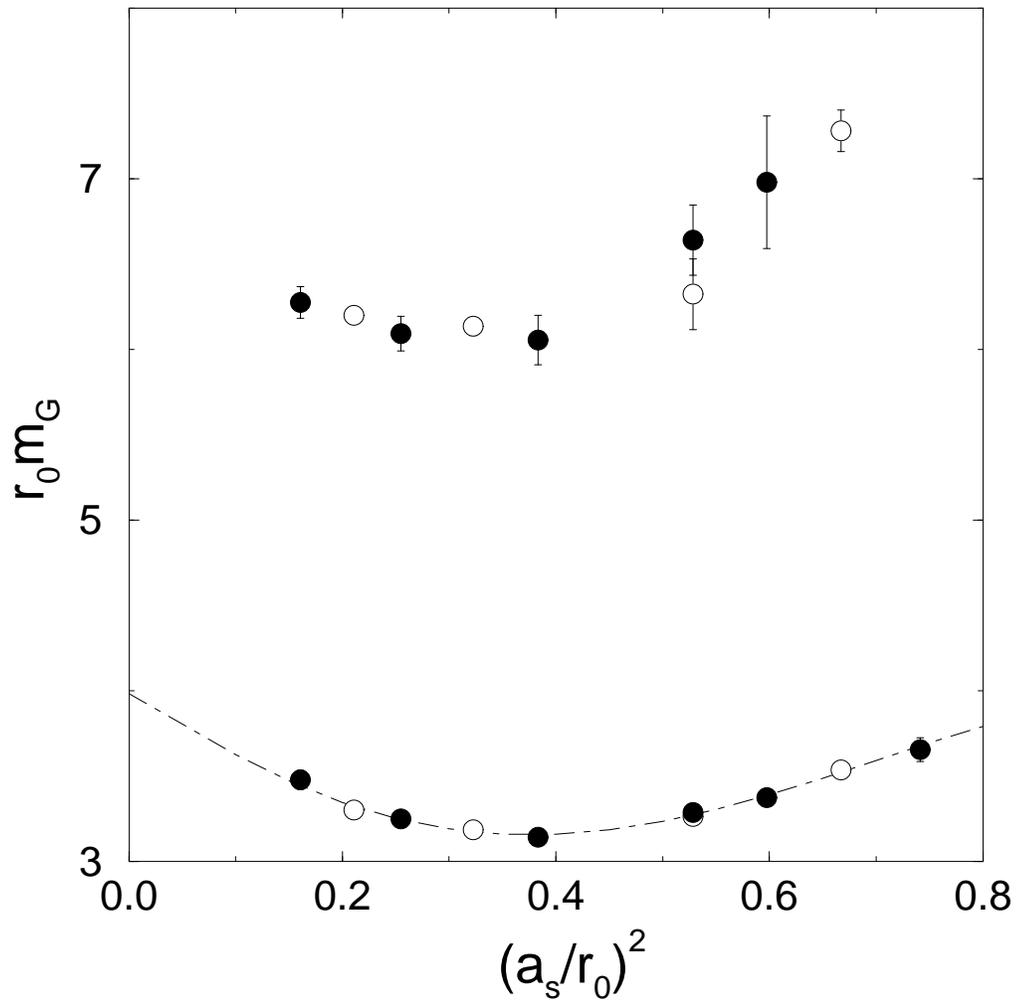}
\end{center}
\caption[figscalar]{
Masses of the scalar glueball and its first-excited state in terms of $r_0$
against the lattice spacing $(a_s/r_0)^2$.  Open and solid symbols
indicate results from the $\xi=5$ and $\xi=3$ simulations, respectively.
The extrapolation to the continuum limit is indicated by
the dash-dotted curve.
\label{fig:scalarscaling}}
\end{figure}

The glueball mass estimates in terms of $a_t^{-1}$ were combined
with the determinations of the hadronic scale $r_0/a_s$.  The results
are shown in Figs.~\ref{fig:Scaling}, \ref{fig:tensor},
and \ref{fig:scalarscaling}.  In these figures, the dimensionless product
of $r_0$ and the glueball mass estimates are shown as functions of
$(a_s/r_0)^2$.  Solid symbols indicate results from the $\xi=3$
simulations, while open symbols are used for the results from the
$\xi=5$ runs.  In Fig.~\ref{fig:Scaling}, the lowest-lying masses
in each of the channels $A_1^{++}$, $E^{++}$, $T_2^{++}$, and $T_1^{+-}$
are compared with results from small-$a_s$ Wilson action simulations.
The lowest-lying and first-excited masses in the $E^{++}$ and $T_2^{++}$
channels are shown in Fig.~\ref{fig:tensor}, and the ground-state
and first-excited state in the $A_1^{++}$ channel are depicted in
Fig.~\ref{fig:scalarscaling}.  To extract physical predictions (for the
pure-gauge theory), the curves in these plots must be extrapolated
to the continuum limit $a_s/r_0\rightarrow 0$.  Discretization errors
are given by the deviations of the finite-$a_s$ results from these
limiting values.

The lowest-lying states in the $E^{++}$ and $T_2^{++}$ channels correspond
to the five polarizations of the tensor $2^{++}$ glueball in the continuum.
Differences between the $E^{++}$ and $T_2^{++}$ masses are a measure of
violations of rotational symmetry due to finite spacing artifacts.  In
Fig.~\ref{fig:Scaling}, such violations are seen to be small for our
less coarse lattices and become appreciable as the spacing gets very large.
Discretization errors in the $T_2^{++}$ exceed those of the $E^{++}$;
on our coarsest lattices, finite spacing errors are only a few per
cent for the $E^{++}$ channel, but about $15\%$ in the $T_2^{++}$ channel.
In the $E^{++}$ channel, the $\xi=3$ results differ very little from
those using the higher aspect ratio $\xi=5$, suggesting that the $O(a_t^2)$
errors are negligible.  However, small differences between the results
from the two anisotropies are visible in the $T_2^{++}$ channel.  One
expects that $O(a_t^2)$ errors will decrease as $\xi$ is increased.
Since the $T_2^{++}$ discretization errors are slightly larger for the
$\xi=5$ runs, $O(a_t^2)$ errors can account for this difference only if
such errors offset the $O(a_s^4)$ errors.

The leading discretization errors in the tensor glueball masses are expected
to be $O(a_t^2,a_s^4,\alpha_s a_s^2)$.  However, we have already argued that
the results in Fig.~\ref{fig:Scaling} imply that the $O(a_t^2)$ errors are
negligible.  Since the action included mean-field correction factors,
we also expected that $O(a_s^4)$ errors would dominate over $O(\alpha_s a_s^2)$
errors and in our continuum limit extrapolations, we assumed that this
was true unless the fit provided compelling evidence to the contrary.
Although we expected the leading discretization errors to be $O(a_s^4)$,
the following three functions were used in our continuum limit
extrapolations:
\begin{eqnarray}
\varphi_0(a_s) &=& r_0 m_G, \label{eq:FitConst}\\
\varphi_2(a_s) &=& r_0 m_G + c_2\ (a_s/r_0)^2,
\label{eq:FitA2} \\
\varphi_4(a_s) &=& r_0 m_G + c_4\ (a_s/r_0)^4,
\label{eq:FitA4}
\end{eqnarray}
where $c_2$, $c_4$ and $r_0 m_G$ are best-fit parameters.
The results of these fits are given in Table~\ref{tab:Extrap-3} for
the $\xi=3$ data and Table~\ref{tab:Extrap-5} for the $\xi=5$ simulations.
Comparing the values of $\chi^2/{\rm dof}$, one sees that
the fitting function $\varphi_4$ was preferred for both the $\xi=3$
and $\xi=5$ results, although only marginally so for $\xi=3$.  Given this
fact and our expectation that $\varphi_4$ should best describe
the leading discretization effects, we took $r_0 m_G$ from the $\varphi_4$
fits as our continuum limit estimates (indicated in boldface in
Tables~\ref{tab:Extrap-3} and \ref{tab:Extrap-5}).  These four estimates
are in very good agreement not only with one another, but also with the
Wilson action estimates.  These fits using $\varphi_4$ are shown
in Fig.~\ref{fig:Scaling}.  For our final estimate of the tensor glueball
mass, we performed a simultaneous fit with the four data sets (two irreps and
two anisotropies) using four separate $\varphi_4$ functions but constraining
the intercept parameter $r_0 m_G$ to be the same for all four fitting
functions.  This yielded $r_0 m(2^{++})=5.85\pm 0.02$ with
$\chi^2/{\rm dof}=1.01$, in agreement with the Wilson action estimate
$r_0 m(2^{++}) = 6.0\pm 0.1$, obtained by fitting all of the Wilson
action measurements shown in Fig.~\ref{fig:Scaling} to $\varphi_0$.

We also examined the discretization errors in the masses of the
first-excited glueball states in the $E^{++}$ and $T_2^{++}$ channels.
These are shown in Fig.~\ref{fig:tensor}.  There are several reasons
for interpreting these data as different polarizations of a
spin-two excited state: the two irreps extrapolate to the same continuum
limit value; if the $T_2^{++}$ state were spin three, then there would be
a degeneracy with the $T_1^{++}$ and $A_2^{++}$ channels
and this was not observed (these results will be presented elsewhere);
if they were polarizations of a spin-four state,
then again, a similar level must also be found in the $T_1^{++}$ channel.
The degeneracy between the two irreps and the weak finite-volume dependence of
their energies also makes an interpretation of this state as a torelon pair 
or a two-scalar-glueball scattering state unlikely, although the mass of
this level lies close to twice the mass of the scalar glueball.
Continuum limit extrapolations were performed using the three 
functions of Eqs.~\ref{eq:FitConst}-\ref{eq:FitA4}; measurements from the two 
largest $a_s$ spacings for $\xi=3$ and the single largest spacing for $\xi=5$ 
were excluded from these fits. The results are given in 
Tables~\ref{tab:Extrap-3} and \ref{tab:Extrap-5}. Again, we expected 
$\varphi_4$ to provide the most reliable extrapolations to the 
$a_s\rightarrow 0$ limit; this was confirmed by the fact that $\varphi_4$ 
yielded $E^{++}$ and $T_2^{++}$ continuum limits in best agreement with each 
other. Differences found between the $\xi=3$ and $\xi=5$ extrapolations were 
not statistically significant. The fits using $\varphi_4$ are shown in 
Fig.~\ref{fig:tensor}. Our final determination of the mass of the excited 
tensor glueball, obtained from a constrained set of four $\varphi_4$, similar 
to the ground-state extraction, was $r_0 m(2^{*++})=8.11\pm 0.04$, where 
$\chi^2/{\rm dof}=2.3$. This mass has not been reliably determined in any 
previous simulations.

Finite spacing errors in the mass of the $T_1^{+-}$ pseudovector
glueball were also studied.  These were found to be small and are shown
in Fig.~\ref{fig:Scaling}.  The results from the different anisotropies are
in good agreement.  Extrapolations to the $a_s\rightarrow 0$ limit were done
using the three functions $\varphi_0$, $\varphi_2$, and $\varphi_4$; the
results of these fits are summarized in Tables~\ref{tab:Extrap-3} and
\ref{tab:Extrap-5}.  The continuum limits obtained from fits to the $\xi=3$
and $\xi=5$ data agreed only for the constant fit form $\varphi_0$.
The fits to $\varphi_2$ and $\varphi_4$ yielded slope parameters ($c_2$ and
$c_4$) with large relative errors and opposite signs for the different
anisotropies.  Hence, the function $\varphi_0$ was used to extrapolate
to the continuum limit.  Due to the very good agreement between the
$\xi=3$ and $\xi=5$ results, all ten data points were used in our
extrapolation fit.  Our estimate from this fit (shown in
Fig.~\ref{fig:Scaling}) was $r_0 m(1^{+-}) =7.21\pm 0.02$, where
$\chi^2/{\rm dof}=1.55$, in agreement with the extrapolation
$r_0 m(1^{+-}) = 7.5\pm 0.4$  using $\varphi_0$ of the Wilson action
results shown in Fig.~\ref{fig:Scaling}.

In contrast to the tensor and pseudovector, the scalar glueball mass
showed significant finite spacing errors (see Fig.~\ref{fig:Scaling}),
even for our less coarse lattices.  As $a_s$ was increased, the
scalar mass first decreased, reached a minimum near $a_s/r_0\sim 0.6$,
then gradually increased.  Near the minimum, the mass was about
$25\%$ lower than estimates of the continuum limit from small-$a_s$
Wilson action computations; a $20\%$ discretization error was observed
in the result from our smallest $a_s$ simulation.  Although the
magnitudes of these errors were significant, they were smaller than
those obtained using the Wilson action by a factor of two.  In order
to extrapolate to the continuum limit, an appropriate fitting function
was needed.  The leading discretization errors were expected to be
$O(a_t^2,a_s^4,\alpha_s a_s^2)$.  However, there were no distinguishable
differences between the $\xi=3$ and $\xi=5$ results, suggesting that
the $O(a_t^2)$ errors were negligible, leaving us to consider
$O(a_s^4, \alpha_s a_s^2)$ effects.  By inspection, one sees that
the fitting form $\varphi_4$, which neglects one-loop $O(\alpha_s a_s^2)$
effects, cannot describe the data, in contrast to the data for the
tensor and pseudovector glueballs.  As $a_s\rightarrow 0$, we expect
the coupling $\alpha_s(a_s)$ to vanish as $-1/\ln(a_s\Lambda)$, where
$\Lambda$ is an appropriate scale parameter.  Hence, we were led to
consider the following four-parameter fitting function:
\begin{equation}
\varphi_{1L}(a_s) = r_0 m_G + c_2\
        \frac{(a_s/r_0)^2}{c_L - \mbox{ln}\left[(a_s/r_0)^2\right]}
        + c_4\ (a_s/r_0)^4. \label{eq:ScalarFit2}
\end{equation}
However, it was not known how reliably the leading perturbative
behavior of $\alpha_s(a_s)$ would describe the true cut-off dependence
of the coupling over the large range of spacings considered.
Taking this into account and inspecting the behavior of the actual data,
we decided to also consider the following simpler quadratic form:
\begin{equation}
\varphi_{2,4}(a_s) = r_0 m_G + c_2\ (a_s/r_0)^2
              + c_4\ (a_s/r_0)^4. \label{eq:ScalarFit1}\\
\end{equation}
Both of these functions were fit to the mass measurements from all ten
simulations; the results of these fits are summarized in
Table~\ref{tab:ScalarExtrap}.  The function $\varphi_{1L}$ yielded
a slightly better fit and a continuum limit for the scalar glueball
mass of $3.98\pm 0.15$.  This fit is shown in Fig.~\ref{fig:Scaling}.
An extrapolation of existing Wilson action data using $\varphi_2$ yielded
$4.33 \pm 0.05$.  Given the quality of the scalar glueball mass estimates
using the Wilson action, this slight discrepancy raises doubts concerning
the reliability of the extrapolation using $\varphi_{1L}$; mass estimates
using the improved action for a few values of $a_s$ smaller than those
considered here would be needed to resolve this discrepancy.

One explanation for the $20\%$ discretization errors in the scalar
glueball mass is that the scalar glueball is extremely small.
However, there is evidence\cite{Heller96,patel} that the
presence of a critical endpoint of a line of phase transitions (not
corresponding to any physical transition found in QCD)
in the fundamental-adjoint coupling plane is responsible for
lowering the scalar glueball mass near the crossover region in the Wilson
action.  It is possible that the scalar glueball mass in the improved action
used here may be similarly influenced.  If so, the fact that this effect
appears to be less pronounced for this action suggests the possible existence
of other perturbatively-improved actions in which the scalar glueball mass
is even less affected by scaling violations.  We are currently
searching for such actions.

Discretization errors in the mass of the first-excited state in the
$A_1^{++}$ channel were also found to be significant, as shown in
Fig.~\ref{fig:scalarscaling}.  The mass of this state is nearly
twice that of the lowest-lying scalar glueball, suggesting that this
state may simply be two glueballs.  Given the significant
discretization errors in the single glueball scaling data, one would expect
similar systematic errors in the two glueball state.  The absence
of any level of similar mass in all other channels justifies the
spin-zero interpretation of this state.  Considering the difficulties
encountered in extrapolating the lowest-lying scalar to the continuum
limit, we made no serious attempt to determine the continuum limit
of this first-excited state.  However, the result of a fit using
$\varphi_{2,4}$ is included in Table~\ref{tab:ScalarExtrap}.  The
possible interpretation of this level as a two glueball system might be
strengthened by a more precise finite-volume study.

\subsection{Conversion to physical units}

In order to convert our glueball mass computations into physical
units, we must specify the value of the hadronic scale.
The hadronic scale $r_0$ has a precise definition in terms of the
static-quark potential.  However, the static-quark potential cannot
be directly measured in an experiment; it must be deduced indirectly
from other observables.  We decided to use a variety of different
physical quantities to deduce $r_0$.

In Table~\ref{hadronictable}, estimates of $r_0^{-1}$ using
the results from various quenched lattice simulations are shown.
For each computation, the quantity used to set the lattice
spacing, such as the mass of the $\rho$ or the $1P-1S$ splitting
in heavy quarkonia, is indicated.  The determination of
$r_0^{-1}$ from $a^{-1}$ was accomplished using values of
$a/r_0$ given in Ref.~\cite{scalebali} for the Wilson gluonic
action at various values of $\beta$, interpolating where necessary.
Note that due to quenching effects, $r_0$ varies with the quantity used
to set the scale.  The entries in the last column of 
Table~\ref{hadronictable} cannot be considered as different measurements
of a single quantity and thus, strictly speaking, their weighted average
has no statistical meaning.  The last column of the table is meant to
illustrate the range of values one obtains for $r_0^{-1}$ when
using various scale setting quantities.  We expect that the value
of $r_0^{-1}$ appropriate for the low-lying glueballs should lie somewhere
within this range.  The simple average $r_0^{-1} = 410\pm 20$~MeV of the
determinations in Table~\ref{hadronictable} was taken as our estimate
of the hadronic scale.

For our final continuum mass estimates of the tensor glueball, we found
$2400\pm 10\pm 120$ MeV (where the first error is statistical and the
second is from uncertainties in the determination of $r_0$).  It is
interesting to note that our mass estimate lies within $8\%$ of the
mass of the $f_J(2220)$ resonance \cite{pdg,glueball2230}, reported to
have quantum numbers $(\mbox{even})^{++}$.  In order to make a direct
comparison with experiment, however, corrections to our result from
light quark effects and mixings with nearby conventional mesons must
be taken into account.  Our estimate of the mass of the first-excited
glueball in the tensor channel was $3320\pm 20\pm 160$ MeV;
for the pseudovector state, we found a mass of $2960 \pm 10 \pm 140$ MeV.
Our estimate from the fit using $\varphi_{1L}$ for the mass of the scalar
glueball was $1630\pm 60\pm 80$~MeV; however, we regard the continuum limit
extrapolation for this state as being less reliable than those for the
other glueballs.

\subsection{Comparison of efficiencies}

A quantitative comparison of efficiencies is difficult to make.
There are many factors which affect the overall efficiency of a
Monte Carlo simulation.  Certainly, the number of link updates is
an important factor.  The speed of an update is, of course,
platform-, action-, and algorithm-dependent.  On the DEC
Alpha-workstations we used, a CM update using an improved action required twice
as much time as for the Wilson action; the improved-action OR updating time was 
three times longer.  Critical slowing down and thermalization are also
contributing factors, but a crucial issue is the reduction of
excited-state contaminations in the glueball correlators.  In our
coarse lattice simulations, we found that current methods for constructing
good glueball operators were very effective in hastening the onset of
plateaux in the effective masses.

Given these difficulties in assessing the efficiency of a glueball
simulation, we decided to make our comparisons based simply on the
number of link updates $N_{lu}$ and the fractional error $\epsilon$ attained
in the final mass estimates.  Since the error in a Monte Carlo estimate
decreases with the number of measurements $N$ as $1/\sqrt{N}$, we expect
that the reciprocal product of the number of link updates and
the square of the fractional error is approximately proportional
to the efficiency of a simulation; we denote this quantity
by ${\cal E}=1/(\epsilon^2 N_{lu})$.  An interesting comparison to make
is between simulations at a small lattice spacing, such as $a_s\sim 0.05$ fm,
using the Wilson action and improved-action simulations at a spacing
$a_s\sim 0.2$ fm.  Such a comparison is relevant because the discretization
errors in these cases are of comparable magnitude, excepting the scalar
glueball mass.  For the $\beta=6.4$
Wilson action run in Ref.~\cite{Chen} by the GF11 collaboration using
gauge invariant glueball operators, a total of $3.13 \times 10^{12}$ link
updates were performed, an error of $2.5\%$ was achieved in the scalar
glueball mass, and a fractional error of $3.6\%$ was obtained for the
tensor mass. Using $r_0=0.48$ fm and $a_s/r_0=0.101(2)$, the lattice volume
in this simulation was $(1.55\ {\rm fm})^2 \times (1.45\ {\rm fm})$.
For the same value of $\beta$ in Ref.~\cite{UKQCD} by the UKQCD
collaboration, $1.35\times 10^{11}$ link updates were made and fractional
errors $3.4\%$, $3.3\%$, and $9\%$ were obtained for the scalar, tensor,
and pseudovector masses, respectively.  The lattice volume was
$(1.55\ {\rm fm} )^3$ for this simulation.  In our $\beta=2.6$, $\xi=3$ run,
$5.76\times 10^9$ link updates were performed and $1.5\%$, $1.0\%$, and
$1.0\%$ errors in the scalar, tensor, and pseudovector masses were achieved.
Our lattice volume was $(1.93\ {\rm fm})^3$ since $a_s/r_0 = 0.4021(9)$.
For our $\beta=2.4$, $\xi=5$ run, the number of link updates was
$9.83\times 10^9$ and the errors obtained in the scalar, tensor, and
pseudovector masses were $1.0\%$, $0.5\%$, and $0.8\%$, respectively.
Using $a_s/r_0 = 0.459(1)$, our lattice volume for this run was
$(1.76\ {\rm fm})^3$.  Thus, the ratios of the ${\cal E}$ values for our
$\beta=2.6$ and $\beta=2.4$ simulations to those of the GF11 run were
$1500$ and $2000$, respectively, for the scalar glueball mass, and
$7000$ and $17000$ for the tensor glueball mass.  The ratios of the
$\beta=2.6$ and $\beta=2.4$ ${\cal E}$-values to those of the UKQCD run
were $120$ and $160$, respectively, for the scalar mass,
$260$ and $600$ for the tensor, and $1900$ and $1700$ for the
pseudovector.

Considering our ten simulations together, a total of $5\times 10^{10}$ link
updates were performed.  For the Wilson action simulations of
Refs.~\cite{Chen,UKQCD}, an estimated $10^{13}$ and $10^{12}$ link updates
were required, respectively, to generate continuum limit results whose
statistical uncertainties were larger (for the tensor and pseudovector states)
than those quoted here: the statistical error on our estimate for
$r_0 m(2^{++})$ was about five times smaller than that from the extrapolation
of the Wilson action results, and for $r_0 m(1^{+-})$, the uncertainty was
twenty times smaller, implying that about $25-400$ times greater statistics
would be required for similar accuracy.  Thus, in total, the anisotropic
lattice simulations were certainly more than 1000 times more efficient.

The above discussion illustrates the computational advantages of extracting
non-scalar glueball masses from simulations on coarse, anisotropic lattices
($a_s\sim 0.2$ fm) using an improved action instead of lattices
for which $a_s\sim 0.05$ fm.  The excellent overlaps achieved from our
variational calculations demonstrate another advantage of simulating on
coarse spatial lattices. The glueball wavefunctions extend over only a few
points of the lattice when $a_s\approx 0.2$~fm.  Thus, variational
calculations using a feasible number of basis functions (a dozen or so)
can yield very good approximations to the glueball wavefunctions.
This will be especially important for future decay constant calculations
and determinations of mixings with non-glueball states.  These
advantages have already enabled us to study the more massive glueball
states which have yet to be simulated reliably using the Wilson action;
these results will be reported elsewhere.

These advantages are less clear for the scalar glueball mass due to
the presence of $20\%$ discretization errors at $a_s\sim 0.2$ fm.
Using the action of Eq.~\ref{eq:ImpAction}, simulations at one or
more lattice spacings smaller than 0.2 fm would be needed to firmly
establish the continuum limit.  A more attractive approach would be
to use an action for which discretization errors in the scalar glueball
mass at $a_s\sim 0.2$ fm are negligibly small.  The search for such
an action is currently underway.

\section{Conclusion}
\label{sec:conclusions}

We have demonstrated the advantages of using anisotropic lattices and an
improved gluonic action for simulating glueballs.  Ten simulations at lattice
spacings ranging from 0.2 to 0.4 fm were performed using DEC 
Alpha-workstations, and the results were extrapolated to the continuum limit.
Results for the masses of the scalar ($0^{++}$), the tensor ($2^{++}$), and the
pseudovector ($1^{+-}$) glueballs in $SU(3)$ pure-gauge theory were presented
in terms of the hadronic scale $r_0^{-1}$.  The continuum limits for the
tensor and pseudovector glueball masses were obtained with uncertainties
of less than $1\%$, significantly improving upon previous estimates from Wilson
action simulations carried out with the aid of super-computer resources. 
Extrapolation of the scalar glueball mass to the continuum limit was
hampered by uncertainties in choosing the fitting function and discretization
errors which were $20-25\%$ even for our smallest lattice spacings; although
uncomfortably large, these finite-spacing errors were half as large as those
obtained using the Wilson action.  Finite volume errors in our results were
shown to be negligible.  The masses of the first excited states in the scalar
and tensor channels were also examined.

Our results show that spatially-coarse, anisotropic lattice simulations
are an effective means of studying gluonic systems. The techniques exploited
here are sufficiently powerful to overcome the difficulties which plague
Monte Carlo calculations involving gluonic excitations.  These methods
should be useful for studying the spectrum of heavier glueball states.
Data for the masses of all twenty lattice irreps of the cubic group (including
parity and charge conjugation) are currently being accumulated in order
to survey the spectrum of $SU(3)$ glueball states below 4 GeV comprehensively.
We shall report on these results in the near future.  We also plan to use the
techniques outlined in this paper to determine various glueball matrix elements
and decay strengths, to investigate the mixings of glueballs with
conventional hadronic states, and to study mesonic states containing
excited glue (the so-called hybrid mesons).  The size of discretization errors
in the scalar glueball mass was the only disappointing aspect of this work; we
are currently investigating a new class of lattice actions with the hope
of reducing these lattice artifacts for $a_s\sim 0.2$ fm to the level
of a few per cent.

\section{Acknowledgments}
We would like to thank Peter Lepage, Julius Kuti, Chris Michael,
Terrence Draper, and Keh-Fei Liu for helpful discussions.
This work was supported by the U.S.~DOE, Grant No.~DE-FG03-90ER40546. MP is
grateful to the University of Kentucky Center for Computational Sciences for
financial support.

%%%%%%%%%%%%%%%%%%%%%%%%%%%%%%%%%%%%%%%%%%%%%%%%%%%%%%%%%%%%%%%
%
%   Illustrations of the glueball operators.
%
%%%%%%%%%%%%%%%%%%%%%%%%%%%%%%%%%%%%%%%%%%%%%%%%%%%%%%%%%%%%%%%

%%%%%%%%%%%%%%%%%%%%%%%%%%%%%%%%%%%%%%%%%%%%%%%%%%%%%%%%%%%%%%%%%%%%%%%%%%
%                    V(2,2,2) effective mass plot                        %
%%%%%%%%%%%%%%%%%%%%%%%%%%%%%%%%%%%%%%%%%%%%%%%%%%%%%%%%%%%%%%%%%%%%%%%%%%

\newpage

%%%%%%%%%%%%%%%%%%%%%%%%%%%%%%%%%%%%%%%%%%%%%%%%%%%%%%%%%%%%%%%%%%%%%%%%%%
%                        Static potential plot                           %
%%%%%%%%%%%%%%%%%%%%%%%%%%%%%%%%%%%%%%%%%%%%%%%%%%%%%%%%%%%%%%%%%%%%%%%%%%

\newpage

%%%%%%%%%%%%%%%%%%%%%%%%%%%%%%%%
%% ANISOTROPY 3:1
%%%%%%%%%%%%%%%%%%%%%%%%%%%%%%%%

%%    BETA = 2.6 GRAPHS    %%%%

%%%%%%%%%%%%%%%%%%%%%%%%%%%%%%%%
%% ANISOTROPY 5:1
%%%%%%%%%%%%%%%%%%%%%%%%%%%%%%%%

%%    BETA = 2.4 GRAPHS    %%%%

%%%%%%%%%%%%%%%%%%%%%%%%%%%%%%%%%%%%%%%%%%%%%%%%%%%%%%%%%%%%%%%%%%%%%%%%%%
%                             SCALING PLOT                               %
%%%%%%%%%%%%%%%%%%%%%%%%%%%%%%%%%%%%%%%%%%%%%%%%%%%%%%%%%%%%%%%%%%%%%%%%%%

%%%%%%%%%%%%%%%%%%%%%%%%%%%%%%%%%%%%%%%%%%%%%%%%%%%%%%%%%%%%%%%%%%%%%%%%%%
%                         TENSOR SCALING PLOT                            %
%%%%%%%%%%%%%%%%%%%%%%%%%%%%%%%%%%%%%%%%%%%%%%%%%%%%%%%%%%%%%%%%%%%%%%%%%%

%%%%%%%%%%%%%%%%%%%%%%%%%%%%%%%%%%%%%%%%%%%%%%%%%%%%%%%%%%%%%%%%%%%%%%%%%%
%                         SCALAR SCALING PLOT                            %
%%%%%%%%%%%%%%%%%%%%%%%%%%%%%%%%%%%%%%%%%%%%%%%%%%%%%%%%%%%%%%%%%%%%%%%%%%

%%%%%%%%%%%%%%%%%%%%%%%%%%%%%%%%%%%%%%%%%%%%%%%%%%%%%%%%%
%                                                       %
%                         TABLES                        %
%                                                       %
%%%%%%%%%%%%%%%%%%%%%%%%%%%%%%%%%%%%%%%%%%%%%%%%%%%%%%%%%

\begin{table}
\caption[tabone]{
 Input parameters used in the glueball simulations.
\label{tab:RunParams}}
\begin{center}
\begin{tabular}{cccl}
$\beta$ & $\xi$ & Lattice &
 \multicolumn{1}{c}{$u_s^4$} \\
\hline
1.7 & 3 & $6^3\times18$  & 0.3075  \\
1.9 & 3 & $6^3\times18$  & 0.340   \\
2.0 & 3 & $8^3\times24$  & 0.356   \\
2.2 & 3 & $8^3\times24$  & 0.3885  \\
2.4 & 3 & $8^3\times24$  & 0.421   \\
2.6 & 3 & $10^3\times30$ & 0.4505  \\
\hline
1.7 & 5 & $6^3\times30$ & 0.295  \\
1.9 & 5 & $6^3\times30$ & 0.328  \\
2.2 & 5 & $8^3\times40$ & 0.378  \\
2.4 & 5 & $8^3\times40$ & 0.409
\end{tabular}
\end{center}
\end{table}

\begin{table}
\caption[potrun]{Various run parameters for the static-quark
potential measurements, including lattice sizes, total
numbers of configurations used, and parameters for the
two different smearing schemes.
\label{potrun}}
\begin{center}
\begin{tabular}{cccccc}
$\beta$ & $\xi$ & Lattice & $\#$ configs &
  $(\lambda_1,n_{\lambda 1})$ & $(\lambda_2, n_{\lambda 2})$
\\ \hline
 1.7 & 3 &  $6^3 \times 18$ & 2275 & $(0.05, 4)$ & $(0.10, 4)$ \\
 1.9 & 3 &  $6^3 \times 18$ & 1280 & $(0.07, 4)$ & $(0.14, 4)$ \\
 2.0 & 3 &  $8^3 \times 24$ &  848 & $(0.07, 4)$ & $(0.15, 4)$ \\
 2.2 & 3 &  $8^3 \times 24$ & 1024 & $(0.10, 6)$ & $(0.16, 6)$ \\
 2.4 & 3 &  $8^3 \times 24$ & 1024 & $(0.12, 6)$ & $(0.25, 6)$ \\
 2.6 & 3 & $10^3 \times 30$ & 1100 & $(0.25, 6)$ & $(0.50, 6)$ \\
 \hline
 1.7 & 5 &  $8^3 \times 40$ &  810 & $(0.10, 6)$ & $(0.25, 6)$ \\
 1.9 & 5 &  $6^3 \times 30$ & 1024 & $(0.08, 4)$ & $(0.16, 4)$ \\
 2.2 & 5 & $12^3 \times 48$ &  315 & $(0.20, 4)$ & $(0.30, 4)$ \\
 2.4 & 5 &  $8^3 \times 40$ &  548 & $(0.20, 6)$ & $(0.40, 6)$
\end{tabular}
\end{center}
\end{table}

\begin{table}
\caption[scale]{Results for the hadronic scale $r_0$ in terms
of the lattice spacing $a_s$.  The Coulombic coupling $e_c$
and the string tension $\sigma$ obtained from a fit of the
on-axis potential to a Coulomb plus linear form
$V(\vec{r})=e_c/r+\sigma r+V_0$ are also given.
\label{r0values}}
\begin{center}
\begin{tabular}{ccllll}
$\beta$ & $\xi$ &
\multicolumn{1}{c}{$r_0/a_s$} &
\multicolumn{1}{c}{$a_s/r_0$} &
\multicolumn{1}{c}{$e_c$} &
\multicolumn{1}{c}{$r_0^2\sigma$} \\ \hline
 1.7 & 3 &  $1.161(2)$ & $0.861(2)$ & $-0.07(1)$ & $1.58(1)$  \\
 1.9 & 3 &  $1.293(3)$ & $0.773(2)$ & $-0.13(2)$ & $1.52(2)$  \\
 2.0 & 3 &  $1.375(1)$ & $0.7271(8)$& $-0.188(7)$& $1.462(7)$ \\
 2.2 & 3 &  $1.615(2)$ & $0.6192(8)$& $-0.288(8)$& $1.362(8)$ \\
 2.4 & 3 &  $1.978(6)$ & $0.505(1)$ & $-0.321(6)$& $1.329(6)$ \\
 2.6 & 3 &  $2.487(5)$ & $0.4021(9)$& $-0.310(2)$& $1.340(2)$ \\
 \hline
 1.7 & 5 &  $1.224(1)$ & $0.8169(9)$& $-0.177(9)$& $1.473(9)$ \\
 1.9 & 5 &  $1.375(2)$ & $0.727(1)$ & $-0.20(1)$ & $1.45(1)$  \\
 2.2 & 5 &  $1.761(2)$ & $0.5680(5)$& $-0.294(4)$& $1.356(4)$ \\
 2.4 & 5 &  $2.180(6)$ & $0.459(1)$ & $-0.308(4)$& $1.342(4)$
\end{tabular}
\end{center}
\end{table}

\begin{table}
\caption[tabemone]{
Results from fits to the $\beta=1.7$, $\xi = 3$ glueball
correlators.  $N_{exp}$ is the number of exponentials used in each
fit, and the fit range refers to the temporal separations
$t_{\rm min}-t_{\rm max}$ used in each fit.  $Q$ is the standard quality of
fit.  Energies are given in $a_t^{-1}$.  Final mass estimates are
highlighted in boldface.
\label{tab:emone}}
\begin{center}
\begin{tabular}{cccccll}
Channel & $N_{exp}$ & fit range & $\chi^2/{\rm dof}$ & $Q$
 & energies & overlaps  \\ \hline
$A_1^{++}$& 1 & $0-5$ &  0.71 & 0.58 & $1.061(7)  $  & $1.000(4)$ \\
          & 1 & $1-5$ &  0.83 & 0.48 & {\bf 1.05(2)} & $0.99(2)$  \\
          & 1 & $2-5$ &  1.20 & 0.30 & $1.04(5)   $  & $0.96(9) $ \\
\hline
$E^{++}$  & 1 & $0-3$ &  0.02 & 0.98 & $1.653(8)  $  & $1.000(3)$ \\
          & 1 & $1-4$ &  0.12 & 0.88 & {\bf 1.65(5)} & $1.00(5)$  \\
          & 2 & $1-3$ &  1.37 & 0.25 & $1.65(5)   $  & $0.98(5) $ \\
          &   &       &       &      & $2.5(2)    $  & $1.1(2)  $ \\
\hline
$T_2^{++}$& 1 & $0-3$ &  0.60 & 0.55 & $1.99(1)   $  & $1.000(2)$ \\
          & 1 & $1-3$ &  0.06 & 0.81 & {\bf 1.92(6)} & $0.94(6)$  \\
          & 2 & $1-3$ &  0.65 & 0.58 & $1.90(6)   $  & $0.88(8) $ \\
          &   &       &       &      & $2.5(3)    $  & $0.8(3)  $ \\
\hline
$T_1^{+-}$& 1 & $1-3$ &  0.62 & 0.43 & {\bf 2.09(9)} & $ 0.95(8)$
\end{tabular}
\end{center}
\end{table}

\begin{table}
\caption[tabemtwo]{
Results from fits to the $\beta=1.9$, $\xi = 3$ glueball
correlators (see Table~\protect\ref{tab:emone}).
\label{tab:emtwo}}
\begin{center}
\begin{tabular}{cccccll}
Channel & $N_{exp}$ & fit range & $\chi^2/{\rm dof}$ & $Q$
 & energies & overlaps  \\ \hline
$A_1^{++}$& 1 & $0-4$ &  1.05 & 0.37 & $0.878(6) $ & $1.000(5)$ \\
          & 1 & $1-4$ &  1.46 & 0.23 &{\bf 0.87(1)}& $0.99(1) $ \\
          & 2 & $1-3$ &  0.74 & 0.53 & $0.88(1)  $ & $1.00(1) $ \\
          &   &       &       &      & $1.8(1)   $ & $1.0(1)  $ \\
\hline
$E^{++}$  & 1 & $0-4$ &  0.11 & 0.95 & $1.493(9) $ & $1.000(3)$ \\
          & 1 & $1-4$ &  0.04 & 0.96 &{\bf 1.47(4)}& $0.98(3) $ \\
          & 2 & $1-3$ &  0.06 & 0.98 & $1.47(4)  $ & $0.95(5) $ \\
          &   &       &       &      & $1.9(1)   $ & $0.91(9) $ \\
\hline
$T_2^{++}$& 1 & $0-3$ &  1.74 & 0.18 & $1.681(8) $ & $1.000(2)$ \\
          & 1 & $1-3$ &  0.22 & 0.64 &{\bf 1.60(4)}& $0.92(4) $ \\
          & 2 & $1-3$ &  0.23 & 0.88 & $1.58(4)  $ & $0.88(4) $ \\
          &   &       &       &      & $2.6(2)   $ & $1.3(3)  $ \\
\hline
$T_1^{+-}$& 1 & $1-3$ &  0.24 & 0.63 &{\bf 1.80(6)}& $0.90(5) $
\end{tabular}
\end{center}
\end{table}

\begin{table}
\caption[tabemthree]{
Results from fits to the $\beta=2.0$, $\xi = 3$ glueball
correlators (see Table~\protect\ref{tab:emone}).
\label{tab:emthree}}
\begin{center}
\begin{tabular}{cccccll}
Channel & $N_{exp}$ & fit range & $\chi^2/{\rm dof}$ & $Q$
 & energies & overlaps  \\ \hline
$A_1^{++}$& 1 & $0-6$ &  0.94 & 0.45 & $0.794(4) $ & $1.000(4)$ \\
          & 1 & $1-6$ &  1.15 & 0.33 &{\bf 0.797(9)}& $1.002(9)$\\
          & 2 & $1-3$ &  2.33 & 0.07 & $0.794(8) $ & $1.001(8)$ \\
          &   &       &       &      & $1.61(5)  $ & $0.95(5) $ \\
\hline
$E^{++}$  & 1 & $0-4$ &  0.67 & 0.57 & $1.423(6) $ & $1.000(2)$ \\
          & 1 & $1-4$ &  0.42 & 0.65 &{\bf 1.40(2)}& $0.97(2) $ \\
          & 2 & $1-3$ &  1.30 & 0.27 & $1.39(2)  $ & $0.96(2) $ \\
          &   &       &       &      & $1.81(6)  $ & $0.94(5) $ \\
\hline
$T_2^{++}$& 1 & $0-4$ &  2.54 & 0.05 & $1.559(5) $ & $1.000(2)$ \\
          & 1 & $1-4$ &  0.67 & 0.51 &{\bf 1.49(2)}& $0.94(2) $ \\
          & 2 & $1-3$ &  0.53 & 0.66 & $1.49(2)  $ & $0.94(2) $ \\
          &   &       &       &      & $2.11(7)  $ & $1.01(7) $ \\
\hline
$T_1^{+-}$& 1 & $1-4$ &  0.02 & 0.98 &{\bf 1.68(3)}& $0.92(3) $
\end{tabular}
\end{center}
\end{table}

\begin{table}
\caption[tabemfour]{
Results from fits to the $\beta=2.2$, $\xi = 3$ glueball
correlators (see Table~\protect\ref{tab:emone}).
\label{tab:emfour}}
\begin{center}
\begin{tabular}{cccccll}
Channel & $N_{exp}$ & fit range & $\chi^2/{\rm dof}$ & $Q$
 & energies & overlaps  \\ \hline
$A_1^{++}$& 1 & $0-7$ &  1.12 & 0.35 & $0.659(4) $ & $0.998(6)$ \\
          & 1 & $1-7$ &  0.76 & 0.58 &{\bf 0.649(8)}& $0.988(8)$\\
          & 2 & $1-4$ &  1.02 & 0.41 & $0.647(8) $ & $0.984(8)$ \\
          &   &       &       &      & $1.25(3)  $ & $0.93(2) $ \\
\hline
$E^{++}$  & 1 & $1-4$ &  0.10 & 0.90 &{\bf 1.19(2)}& $0.95(2) $ \\
          & 1 & $2-4$ &  0.10 & 0.75 & $1.17(6)  $ & $0.9(1)  $ \\
          & 2 & $1-3$ &  0.32 & 0.81 & $1.19(2)  $ & $0.95(2) $ \\
          &   &       &       &      & $1.62(4)  $ & $0.99(4) $ \\
\hline
$T_2^{++}$& 1 & $0-4$ &  2.32 & 0.07 & $1.280(6) $ & $1.001(2)$ \\
          & 1 & $1-4$ &  0.92 & 0.40 &{\bf 1.24(2)}& $0.96(2) $ \\
          & 1 & $2-4$ &  0.02 & 0.88 & $1.16(6)  $ & $0.8(1)  $ \\
          & 2 & $1-3$ &  1.30 & 0.27 & $1.24(2)  $ & $0.96(2) $ \\
          &   &       &       &      & $1.65(5)  $ & $0.87(4) $ \\
\hline
$T_1^{+-}$& 1 & $1-4$ &  1.13 & 0.32 &{\bf 1.48(3)}& $0.98(3) $
\end{tabular}
\end{center}
\end{table}

\begin{table}
\caption[tabemfive]{
Results from fits to the $\beta=2.4$, $\xi = 3$ glueball
correlators (see Table~\protect\ref{tab:emone}).
\label{tab:emfive}}
\begin{center}
\begin{tabular}{cccccll}
Channel & $N_{exp}$ & fit range & $\chi^2/{\rm dof}$ & $Q$
 & energies & overlaps  \\ \hline
$A_1^{++}$& 1 & $1-8$ &  0.19 & 0.98 &{\bf 0.548(6)}& $0.988(6)$ \\
          & 1 & $2-8$ &  0.22 & 0.96 & $0.550(9) $ & $0.99(2)  $ \\
          & 2 & $1-4$ &  1.12 & 0.35 & $0.550(6) $ & $0.991(6) $ \\
          &   &       &       &      & $1.03(2)  $ & $0.96(1)  $ \\
\hline
$E^{++}$  & 1 & $0-5$ &  1.30 & 0.27 & $1.012(4) $ & $1.000(3) $ \\
          & 1 & $1-5$ &  0.31 & 0.82 &{\bf 0.995(9)}& $0.982(9)$ \\
          & 2 & $1-3$ &  2.14 & 0.09 & $0.993(8) $ & $0.982(8) $ \\
          &   &       &       &      & $1.37(2)  $ & $1.00(2)  $ \\
\hline
$T_2^{++}$& 1 & $0-5$ &  3.98 & 0.00 & $1.035(4) $ & $1.001(2) $ \\
          & 1 & $1-5$ &  0.28 & 0.84 &{\bf 1.006(8)}& $0.969(9)$ \\
          & 2 & $1-3$ &  0.53 & 0.66 & $1.006(8) $ & $0.966(8) $ \\
          &   &       &       &      & $1.42(2)  $ & $0.99(2)  $ \\
\hline
$T_1^{+-}$& 1 & $1-4$ &  0.75 & 0.47 &{\bf 1.24(1)}& $0.98(1)  $
\end{tabular}
\end{center}
\end{table}

\begin{table}
\caption[tabemsix]{
Results from fits to the $\beta=2.6$, $\xi = 3$ glueball
correlators (see Table~\protect\ref{tab:emone}).
\label{tab:emsix}}
\begin{center}
\begin{tabular}{cccccll}
Channel & $N_{exp}$ & fit range & $\chi^2/{\rm dof}$ & $Q$
 & energies & overlaps  \\ \hline
$A_1^{++}$& 1 & $1-10$&  0.88 & 0.53 &{\bf 0.464(7)}& $0.986(8) $ \\
          & 1 & $2-10$&  1.01 & 0.42 & $0.46(1)   $ & $0.98(2)  $ \\
          & 2 & $1-4$ &  0.89 & 0.50 & $0.464(6)  $ & $0.988(8) $ \\
          &   &       &       &      & $0.84(1)   $ & $0.96(1)  $ \\
\hline
$E^{++}$  & 1 & $1-6$ &  0.60 & 0.67 &{\bf 0.781(8)}& $0.984(8) $ \\
          & 1 & $2-6$ &  0.36 & 0.78 & $0.76(2)   $ & $0.94(4)  $ \\
          & 2 & $1-4$ &  0.89 & 0.50 & $0.782(9)  $ & $0.982(9) $ \\
          &   &       &       &      & $1.09(2)   $ & $0.95(2)  $ \\
\hline
$T_2^{++}$& 1 & $1-6$ &  0.31 & 0.87 &{\bf 0.777(8)}& $0.977(8) $ \\
          & 1 & $2-6$ &  0.12 & 0.95 & $0.76(2)   $ & $0.95(3)  $ \\
          & 2 & $1-4$ &  0.53 & 0.79 & $0.777(8)  $ & $0.976(8) $ \\
          &   &       &       &      & $1.12(1)   $ & $0.99(1)  $ \\
\hline
$T_1^{+-}$& 1 & $1-5$ &  0.27 & 0.85 &{\bf 0.97(1)} & $0.97(1)  $
\end{tabular}
\end{center}
\end{table}

\begin{table}
\caption[tabemseven]{
Results from fits to the $\beta=1.7$, $\xi = 5$ glueball
correlators (see Table~\protect\ref{tab:emone}).
\label{tab:emseven}}
\begin{center}
\begin{tabular}{cccccll}
Channel & $N_{exp}$ & fit range & $\chi^2/{\rm dof}$ & $Q$
 & energies & overlaps  \\ \hline
$A_1^{++}$& 1 & $0-5$ &  1.36 & 0.24 & $0.585(3)$ & $0.999(4)$  \\
          & 1 & $1-5$ &  0.26 & 0.86 &{\bf 0.578(5)}& $0.992(5)$\\
          & 2 & $1-4$ &  0.38 & 0.89 & $0.578(5)$ & $0.992(5)$  \\
          &   &       &       &      & $1.19(2) $ & $0.97(2) $  \\
\hline
$E^{++}$  & 1 & $0-5$ &  2.09 & 0.08 & $0.943(3)$ & $1.000(2)$  \\
          & 1 & $1-5$ &  0.12 & 0.95 &{\bf 0.924(8)}& $0.981(7)$\\
          & 2 & $1-4$ &  0.29 & 0.94 & $0.924(8)$ & $0.979(7)$  \\
          &   &       &       &      & $1.29(2) $ & $0.98(1) $  \\
\hline
$T_2^{++}$& 1 & $0-5$ &  1.25 & 0.29 & $1.107(3)$ & $1.001(2)$  \\
          & 1 & $1-5$ &  1.58 & 0.19 &{\bf 1.103(8)}& $0.997(9)$\\
          & 2 & $1-4$ &  0.87 & 0.52 & $1.104(9)$ & $0.997(9)$  \\
          &   &       &       &      & $1.41(2) $ & $0.94(2) $  \\
\hline
$T_1^{+-}$& 1 & $0-3$ &  2.65 & 0.07 & $1.214(4)$ & $1.000(2)$  \\
          & 1 & $1-3$ &  0.21 & 0.65 &{\bf 1.19(1)}& $0.97(1) $ \\
          & 2 & $1-3$ &  0.39 & 0.76 & $1.18(1) $ & $0.97(1) $  \\
          &   &       &       &      & $1.55(3) $ & $0.92(2) $
\end{tabular}
\end{center}
\end{table}

\begin{table}
\caption[tabemeight]{
Results from fits to the $\beta=1.9$, $\xi = 5$ glueball
correlators (see Table~\protect\ref{tab:emone}).
\label{tab:emeight}}
\begin{center}
\begin{tabular}{cccccll}
Channel & $N_{exp}$ & fit range & $\chi^2/{\rm dof}$ & $Q$
 & energies & overlaps  \\ \hline
$A_1^{++}$& 1 & $1-9$ &  1.26 & 0.26 &{\bf 0.475(4)}&$0.992(5)$ \\
          & 1 & $2-9$ &  1.11 & 0.35 & $0.468(6) $ & $0.98(1) $ \\
          & 2 & $2-4$ &  1.03 & 0.38 & $0.468(6) $ & $0.98(1) $ \\
          &   &       &       &      & $0.92(3)  $ & $0.85(5) $ \\
\hline
$E^{++}$  & 1 & $1-6$ &  0.42 & 0.80 &{\bf 0.844(6)}& $0.992(6)$ \\
          & 1 & $2-6$ &  0.20 & 0.90 & $0.83(1)   $ & $0.97(2) $ \\
          & 2 & $1-4$ &  0.69 & 0.66 & $0.844(6)  $ & $0.992(6)$ \\
          &   &       &       &      & $1.09(1)   $ & $0.95(1) $ \\
\hline
$T_2^{++}$& 1 & $1-5$ &  0.91 & 0.43 &{\bf 0.918(7)}& $0.982(6)$ \\
          & 1 & $2-5$ &  1.11 & 0.33 & $0.91(2)   $ & $0.96(3) $ \\
          & 2 & $1-4$ &  1.44 & 0.19 & $0.918(6)  $ & $0.981(6)$ \\
          &   &       &       &      & $1.23(1)   $ & $0.96(1) $ \\
\hline
$T_1^{+-}$& 1 & $1-5$ &  0.30 & 0.83 &{\bf 1.053(8)}& $0.979(7)$ \\
          & 1 & $2-5$ &  0.18 & 0.84 & $1.04(2)   $ & $0.95(4) $ \\
          & 2 & $1-4$ &  0.48 & 0.82 & $1.052(9)  $ & $0.97(1) $ \\
          &   &       &       &      & $1.30(1)   $ & $0.92(1) $
\end{tabular}
\end{center}
\end{table}

\begin{table}
\caption[tabemnine]{
Results from fits to the $\beta=2.2$, $\xi = 5$ glueball
correlators (see Table~\protect\ref{tab:emone}).
\label{tab:emnine}}
\begin{center}
\begin{tabular}{cccccll}
Channel & $N_{exp}$ & fit range & $\chi^2/{\rm dof}$ & $Q$
 & energies & overlaps  \\ \hline
$A_1^{++}$& 1 & $1-14$ & 0.60 & 0.84 &{\bf 0.362(3)}& $0.998(5)$ \\
          & 1 & $2-14$ & 0.50 & 0.90 & $0.366(4) $  & $1.004(7)$ \\
          & 2 & $1-4$ &  0.86 & 0.52 & $0.362(3)  $ & $0.998(5)$ \\
          &   &       &       &      & $0.697(6)  $ & $0.970(7)$ \\
\hline
$E^{++}$  & 1 & $1-7$ &  1.33 & 0.25 &{\bf 0.667(4)}& $0.982(4)$ \\
          & 1 & $2-7$ &  1.65 & 0.16 & $0.666(7) $  & $0.98(1) $ \\
          & 2 & $1-4$ &  0.57 & 0.75 & $0.667(4) $  & $0.980(5)$ \\
          &   &       &       &      & $0.878(7) $  & $0.968(7)$ \\
\hline
$T_2^{++}$& 1 & $1-8$ &  1.05 & 0.39 &{\bf 0.686(4)}& $0.983(3)$ \\
          & 1 & $2-8$ &  1.20 & 0.31 & $0.683(6) $  & $0.98(1) $ \\
          & 2 & $1-4$ &  0.77 & 0.59 & $0.686(3) $  & $0.982(3)$ \\
          &   &       &       &      & $0.938(5) $  & $0.970(5)$ \\
\hline
$T_1^{+-}$& 1 & $1-6$ &  0.48 & 0.75 &{\bf 0.819(4)}& $0.974(4)$ \\
          & 1 & $2-6$ &  0.57 & 0.63 & $0.82(1)  $  & $0.98(2) $ \\
          & 2 & $1-4$ &  0.60 & 0.73 & $0.820(4) $  & $0.974(4)$ \\
          &   &       &       &      & $1.025(8) $  & $0.956(7)$
\end{tabular}
\end{center}
\end{table}

\begin{table}
\caption[tabemten]{
Results from fits to the $\beta=2.4$, $\xi = 5$ glueball
correlators (see Table~\protect\ref{tab:emone}).
\label{tab:emten}}
\begin{center}
\begin{tabular}{cccccll}
Channel & $N_{exp}$ & fit range & $\chi^2/{\rm dof}$ & $Q$
 & energies & overlaps  \\ \hline
$A_1^{++}$& 1 & $1-13$&  1.35 & 0.19 &{\bf 0.303(3)}& $0.995(7) $ \\
          & 1 & $2-13$&  1.15 & 0.32 & $0.307(4)  $ & $1.000(8) $ \\
          & 2 & $1-5$ &  0.86 & 0.56 & $0.304(3)  $ & $0.994(7) $ \\
          &   &       &       &      & $0.569(4)  $ & $0.972(5) $ \\
\hline
$E^{++}$  & 1 & $1-9$ &  1.33 & 0.23 &{\bf 0.538(3)}& $0.992(3) $ \\
          & 1 & $2-9$ &  1.46 & 0.19 & $0.536(5)  $ & $0.986(9) $ \\
          & 2 & $2-5$ &  2.71 & 0.01 & $0.538(5)  $ & $0.99(1)  $ \\
          &   &       &       &      & $0.723(9)  $ & $0.94(1)  $ \\
\hline
$T_2^{++}$& 1 & $1-9$ &  1.08 & 0.38 &{\bf 0.542(2)}& $0.988(3) $ \\
          & 1 & $2-7$ &  0.34 & 0.85 & $0.540(4)  $ & $0.982(7) $ \\
          & 2 & $2-5$ &  1.28 & 0.26 & $0.538(4)  $ & $0.978(8) $ \\
          &   &       &       &      & $0.730(8)  $ & $0.94(1)  $ \\
\hline
$T_1^{+-}$& 1 & $2-6$ &  1.03 & 0.38 &{\bf 0.652(5)}& $0.95(1)  $ \\
          & 2 & $2-6$ &  1.77 & 0.07 & $0.648(6)  $ & $0.95(1)  $ \\
          &   &       &       &      & $0.794(9)  $ & $0.88(2)  $
\end{tabular}
\end{center}
\end{table}

%%%%%%%%%%%%%%%%%%%%%%%%%%%%%%%%%%%%%%%%%%%%%%%%%%%%%%%%%%%%%%%%%%%%%%%%%%
%                         FINAL MASSES ; SUMMARY                         %
%%%%%%%%%%%%%%%%%%%%%%%%%%%%%%%%%%%%%%%%%%%%%%%%%%%%%%%%%%%%%%%%%%%%%%%%%%

\begin{table}
\caption[tabsix]{
Summary of final mass estimates from all $\xi=3$ and $\xi=5$
simulations.
\label{tab:final}}
\begin{center}
\begin{tabular}{ccllll}
$\beta$ & $\xi$ & $a_tm(A_1^{++})$ & $a_tm(E^{++})$ & $a_tm(T_2^{++})$
& $a_tm(T_1^{+-})$ \\  \hline
1.7 & 3 & 1.05(2)  & 1.65(5)  & 1.92(6)  & 2.09(9)  \\
1.9 & 3 & 0.87(1)  & 1.47(4)  & 1.60(4)  & 1.80(6)  \\
2.0 & 3 & 0.797(9) & 1.40(2)  & 1.49(2)  & 1.68(3)  \\
2.2 & 3 & 0.649(8) & 1.19(2)  & 1.24(2)  & 1.48(3)  \\
2.4 & 3 & 0.548(6) & 0.995(9) & 1.006(8) & 1.24(1)  \\
2.6 & 3 & 0.464(7) & 0.781(8) & 0.777(8) & 0.97(1)  \\
\hline
1.7 & 5 & 0.578(5) & 0.924(8) & 1.103(8) & 1.19(1)  \\
1.9 & 5 & 0.475(4) & 0.844(6) & 0.918(7) & 1.053(8) \\
2.2 & 5 & 0.362(3) & 0.667(4) & 0.686(4) & 0.819(4) \\
2.4 & 5 & 0.303(3) & 0.538(3) & 0.542(2) & 0.652(5)
\end{tabular}
\end{center}
\end{table}

%%%%%%%%%%%%%%%%%%%%%%%%%%%%%%%%%%%%%%%%%%%%%%%%%%%%%%%%%%%%%%%%%%%%%%%%%%
%                          FINITE VOLUME MASSES                          %
%%%%%%%%%%%%%%%%%%%%%%%%%%%%%%%%%%%%%%%%%%%%%%%%%%%%%%%%%%%%%%%%%%%%%%%%%%

\begin{table}
\caption[tabFV]{
Glueball mass estimates in terms of $a_t^{-1}$ for $\beta=2.4$,
$\xi=3$ and various lattice volumes.
\label{tab:FinVolVals}}
\begin{center}
\begin{tabular}{cllll}
Channel     & $L_s/a_s=4$& $L_s/a_s=5$  &  $L_s/a_s=6$  & $L_s/a_s=8$ \\
 \hline
$A_1^{++}$  &   0.483(6) & 0.551(6) &  0.545(8)  & 0.548(6)  \\
$A_1^{*++}$ &   0.863(9) & 0.97(1)  &  1.00(3)   & 1.027(17) \\
$E^{++}$    &   0.848(9) & 0.980(10)&  1.004(19) & 0.995(9)  \\
$E^{*++}$   &   1.11(2)  & 1.27(2)  &  1.41(4)   & 1.37(2)   \\
$T2^{++}$   &   1.032(9) & 1.005(9) &  1.002(14) & 1.006(8)  \\
$T2^{*++}$  &   1.37(2)  & 1.42(2)  &  1.40(3)   & 1.42(2)   \\
$T1^{+-}$   &   1.245(13)& 1.235(11)&  1.209(25) & 1.237(14)
\end{tabular}
\end{center}
\end{table}

%%%%%%%%%%%%%%%%%%%%%%%%%%%%%%%%%%%%%%%%%%%%%%%%%%%%%%%%%%%%%%%%%%%%%%%%%%
%                        FINITE VOLUME EFFECTS                           %
%%%%%%%%%%%%%%%%%%%%%%%%%%%%%%%%%%%%%%%%%%%%%%%%%%%%%%%%%%%%%%%%%%%%%%%%%%

\begin{table}
\caption[tabtwo]{
 The effects of simulating in a finite box: results from fits of
 Eq.~\protect{\ref{eq:LuscherFinVol}} to the energy estimates given
 in Table~\protect{\ref{tab:FinVolVals}} from lattices
 of spatial extent $L_s/a_s = 4,5,6,8$.  The $T_1^{+-}$ fit also
 includes an energy estimate for $L_s/a_s=3$.  The final column estimates
 the expected finite-volume errors in glueball masses from the $L_s/a_s = 8$
 simulation at $\beta=2.4$ and $\xi=3$.  These errors are estimated
 by $m_G(8\omega\xi)/m_G(\infty)-1$ using
 Eq.~\protect{\ref{eq:LuscherFinVol}}.
\label{tab:FinVolFits}}
\begin{center}
\begin{tabular}{clcll}
Channel   & $a_t m_G(\infty)$ & $\lambda_G$    & $\chi^2/$dof &
 \% correction \\ \hline
$A_1^{++}$& 0.554(4)  & $260(37)$ & 2.3    & $-0.020$ \\
$E^{++}  $& 1.002(7)  & $319(23)$ & 0.62   & $-0.024$ \\
$T_2^{++}$& 1.003(6)  & $-59(24)$ & 0.15   & $+0.004$ \\
$T_1^{+-}$& 1.223(7)  & $-66(4)$  & 1.05   & $+0.005$
\end{tabular}
\end{center}
\end{table}

%%%%%%%%%%%%%%%%%%%%%%%%%%%%%%%%%%%%%%%%%%%%%%%%%%%%%%%%%%%%%%%%%%%%%%%%%%
%                      CONTINUUM EXTRAPOLATIONS                          %
%%%%%%%%%%%%%%%%%%%%%%%%%%%%%%%%%%%%%%%%%%%%%%%%%%%%%%%%%%%%%%%%%%%%%%%%%%

\begin{table}
\caption[tabthree]{
 Extrapolations of the glueball mass estimates to the continuum limit
 for the $\xi=3$ runs.  The three scaling forms $\varphi_0$, $\varphi_2$,
 and $\varphi_4$ which are fit to the data are given in
 Eqs.~\protect{\ref{eq:FitConst}}, \protect{\ref{eq:FitA2}} and
 \protect{\ref{eq:FitA4}}. The values indicated in bold are taken as our
 final continuum mass estimates.
\label{tab:Extrap-3} }
\begin{center}
\begin{tabular}{cclccc}
 Channel  & fit function & $r_0 m_G$    & $c_2 $    & $c_4 $    &
 $\chi^2/$dof \\ \hline
$E^{++}$  & $\varphi_0$   & 5.83(3)      &     ---    &    ---    & 0.71  \\
          & $\varphi_2$   & 5.91(7)      & $-0.26(20)$&    ---    & 0.49  \\
          & $\varphi_4$   &{\bf 5.87(5)} &     ---    &$-0.33(25)$& 0.46  \\
\hline
$T_2^{++}$& $\varphi_0$   & 5.98(3)      &     ---    &    ---    & 5.35  \\
          & $\varphi_2$   & 5.66(7)      & 1.02(21)   &    ---    & 0.90  \\
          & $\varphi_4$   &{\bf 5.83(5)} &     ---    &   1.29(27)& 0.87  \\
\hline
$T_1^{+-}$& $\varphi_0$   &{\bf 7.22(5)} &     ---    &    ---    & 1.81  \\
          & $\varphi_2$   & 7.44(11)     &$-0.71(31)$ &    ---    & 0.96  \\
          & $\varphi_4$   & 7.32(7)      &     ---    &$-0.87(40)$& 1.07  \\
\hline
$E^{*++}$ & $\varphi_0$   & 7.99(8)      &     ---    &    ---    & 2.44  \\
          & $\varphi_2$   & 8.52(20)     & $-1.8(6)$  &    ---    & 0.31  \\
          & $\varphi_4$   &{\bf 8.25(12)}&     ---    & $-2.6(9)$ & 0.12  \\
\hline
$T_2^{*++}$& $\varphi_0$  & 8.37(6)      &     ---    &    ---    & 1.33  \\
           & $\varphi_2$  & 8.32(16)     &  0.2(7)    &    ---    & 1.94  \\
           & $\varphi_4$  &{\bf 8.34(8) }&     ---    &   0.5(11) & 1.89
\end{tabular}
\end{center}
\end{table}

\begin{table}
\caption[tabfour]{
 Extrapolations of the glueball mass estimates to the continuum limit
 for the $\xi=5$ runs.  The three scaling forms $\varphi_0$, $\varphi_2$,
 and $\varphi_4$ which are fit to the data are given in
 Eqs.~\protect{\ref{eq:FitConst}}, \protect{\ref{eq:FitA2}} and
 \protect{\ref{eq:FitA4}}. The values indicated in bold are taken as our
 final continuum mass estimates.
\label{tab:Extrap-5} }
\begin{center}
\begin{tabular}{cclccc}
 Channel   & fit function & $r_0 m_G$   & $c_2 $    & $c_4 $    &
$\chi^2/$dof\\ \hline
$E^{++}$   & $\varphi_0$  & 5.82(2)     &     ---   &     ---    & 5.03 \\
           & $\varphi_2$  & 5.98(5)     &$-0.41(12)$&     ---    & 1.55 \\
           & $\varphi_4$  &{\bf 5.91(3)}&     ---   & $-0.50(14)$& 0.81 \\
\hline
$T_2^{++}$ & $\varphi_0$  & 6.12(2)     &     ---   &     ---    & 82   \\
           & $\varphi_2$  & 5.53(4)     & 1.69(11)  &     ---    & 5.16 \\
           & $\varphi_4$  &{\bf 5.82(3)}&     ---   &   2.02(13) & 1.32 \\
\hline
$T_1^{+-}$ & $\varphi_0$  &{\bf 7.21(2)}&     ---   &     ---    & 1.61 \\
           & $\varphi_2$  & 7.09(7)     &  0.31(16) &     ---    & 0.57 \\
           & $\varphi_4$  & 7.15(4)     &     ---   &   0.32(18) & 0.80 \\
\hline
$E^{*++}$  & $\varphi_0$  & 7.67(4)     &     ---   &     ---    & 5.88 \\
           & $\varphi_2$  & 8.12(14)    &$-1.2(4)$  &     ---    & 0.02 \\
           & $\varphi_4$  &{\bf 7.91(8)}&     ---   & $-1.5(4)$  & 0.27 \\
\hline
$T_2^{*++}$& $\varphi_0$  & 8.29(3)     &     ---   &     ---    & 9.17 \\
           & $\varphi_2$  & 7.80(10)    & 1.3(3)    &     ---    & 1.37 \\
           & $\varphi_4$  &{\bf 8.06(6)}&     ---   &    1.4(3)  & 2.48
\end{tabular}
\end{center}
\end{table}

%%%%%%%%%%%%%%%%%%%%%%%%%%%%%%%%%%%%%%%%%%%%%%%%%%%%%%%%%%%%%%%%%%%%%%%%%%

\begin{table}
\caption[tabfive]{
 Extrapolations of the scalar glueball mass estimates to the continuum limit.
 The fit functions used are given in Eqs.~\protect{\ref{eq:ScalarFit2}} and
 \protect{\ref{eq:ScalarFit1}}. The value indicated in bold is our final
 continuum mass estimate for the scalar glueball.
\label{tab:ScalarExtrap} }
\begin{center}
\begin{tabular}{cclcccc}
State   & fit function &
$r_0 m_G$ & $c_2 $ & $c_4 $ & $c_L$& $\chi^2/$dof   \\
\hline
$A_1^{++}$
 & $\varphi_{1L}$  &{\bf 3.98(15)}& $-18(4)$  & 18(5)  & 0.96(13)& 0.25 \\
 & $\varphi_{2,4}$ &      3.86(8) & $-3.5(4)$ & 4.5(5) &  ---    & 0.55 \\
 \hline
$A_1^{*++}$
 & $\varphi_{2,4}$ & 6.93(19) &$-5.3(1.1)$& 8.8(1.4)& ---  & 0.41
\end{tabular}
\end{center}
\end{table}

%%%%%%%%%%%%%%%%%%%%%%%%%%%%%%%%%%%%%%%%%%%%%%%%%%%%%%%%%%%%%%%%%%%%%%%%%%

\begin{table}
\caption[hadronic]{
Estimates of $r_0^{-1}$ using results from various quenched
lattice simulations with the Wilson gluonic action.  The simple
average $r_0^{-1}=410(20)$ MeV of the last column is taken as
our estimate.
\label{hadronictable}}
\begin{center}
\begin{tabular}{clccll}
Source & $\beta$ & Quark Action & Scale Setting Quantity & $a^{-1}$ (GeV)
 & $r_0^{-1}$ (MeV) \\ \hline
NRQCD\cite{nrqcd}  &  6.0  &  NRQCD &  $\Upsilon(2S\!-\!1S,1P\!-\!1S)$
  & 2.4(1) & 434(23)\\
NRQCD\cite{nrqcdc}  &  5.7  &  NRQCD &  $J/\psi(1P\!-\!1S)$
  &  1.23(4)  & 430(16)\\
LANL\cite{tanmoy}   &  6.0  &  Wilson & $M_\rho$
  & 2.330(41) & 422(16) \\
GF11\cite{weingarten}   &  6.17 &  Wilson & $M_\phi$
  &  2.93(11) & 419(17) \\
JLQCD\cite{jlqcd}  &  6.1  &  Wilson & $J/\psi(1P\!-\!1S)$
  & 2.54(7) & 394(13) \\
JLQCD\cite{jlqcd}  &  6.3  &  Wilson & $J/\psi(1P\!-\!1S)$
  & 3.36(11) & 401(14)\\
JLQCD\cite{jlqcd2} &  6.3  &  Wilson & $M_\rho$
  & 3.41(20) & 406(24)\\
FNAL\cite{fermilab}   &  6.1  &  FNAL  & $J/\psi(1P\!-\!1S)$
  & 2.55(8) & 395(14) \\
BLS\cite{bls}    &  6.3  &  heavy-light & $f_\pi$
  & 3.21(9) & 383(11) \\
BLS\cite{bls}    &  6.3  &  heavy-light & $M_\rho$
  & 3.44(9) & 410(11) \\
UKQCD\cite{allton}  &  6.2  &  Wilson & $M_\rho$
  & 2.77(16) & 376(22)
\end{tabular}
\end{center}
\end{table}

%%%%%%%%%%%%%%%%%%%%%%%%%%%%%%%%%%%%%%%%%%%%%%%%%%%%%%%%%%%%%%%%%%%%%%%%%%

\end{document}